\documentclass[]{article}

\usepackage{amsfonts}
\usepackage{url}
\usepackage{amsmath}
\usepackage{graphics}
\usepackage{color}

\newtheorem{defn}{Definition}
\newtheorem{theorem}{Theorem}
\newtheorem{prop}{Proposition}
\newtheorem{lemma}{Lemma}
\newtheorem{cor}{Corollary}
\newcommand{\QED}{\ensuremath{\hfill \Box}}
\newcommand{\sgn}{\ensuremath{\mathrm{sgn}}}
\newcommand{\markov}{\ensuremath{\leftrightarrow}}

\begin{document}

\title{An Infeasibility Result for the \\
Multiterminal Source-Coding Problem%
\footnote{This research was supported by DARPA under Grants~F30602-00-2-0538
and~N66001-00-C-8062, under Grant~N00014-1-0637 from the Office of
Naval Research, and under Grant~ECS-0123512
from the National Science Foundation.}}

\author{Aaron B.\ Wagner\thanks{Coordinated Science Laboratory,
University of Illinois at Urbana-Champaign and
School of Electrical and Computer Engineering, Cornell
University. Email: \texttt{wagner@ece.cornell.edu}.} \ and
Venkat Anantharam\thanks{Department of Electrical Engineering
and Computer Sciences,
University of California, Berkeley.
Email: \texttt{ananth@eecs.berkeley.edu}.}
}

\date{November 29, 2005}

\maketitle

\begin{abstract}
We prove a new outer bound on the rate-distortion region for the multiterminal
source-coding problem. This bound subsumes the best outer bound
in the literature and improves upon it strictly in some cases.
The improved bound enables us to obtain a new,
conclusive result for the binary erasure version of the ``CEO problem.''
The bound recovers many of the converse results that
have been established for special cases of the problem, including
the recent one for the Gaussian version of the CEO
problem.
\end{abstract}

\section{Introduction}

\begin{figure}
\begin{center}
\scalebox{.9}{\input{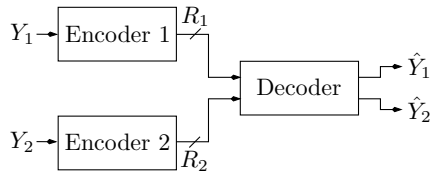}}
\end{center}
\caption{Separate encoding of correlated sources.}
\label{MTSC:basic:fig}
\end{figure}

In their lauded paper~\cite{Slepian:SW}, 
David Slepian and Jack~K.~Wolf
characterize the information
rates needed to losslessly communicate two
correlated, memoryless information sources when 
these sources are encoded separately.
Their well-known
result states that two discrete sources
$Y_1$ and $Y_2$ can be losslessly reproduced
if
\begin{align*}
R_1 & > H(Y_1|Y_2) \\
R_2 & > H(Y_2|Y_1) \\
R_1 + R_2 & > H(Y_1,Y_2),
\end{align*}
where $R_1$ is the rate of the encoder observing
$Y_1$ and $R_2$ is the rate of the encoder observing
$Y_2$. Conversely, lossless reproduction is not possible
if $(R_1,R_2)$ lies outside the closure
of this region.
See Cover and Thomas~\cite[Section~14.4]{Cover:IT} or 
Csisz\'{a}r and K\"{o}rner~\cite[Section~3.1]{CK:IT} for precise
statements of the result and modern proofs.
This result is naturally viewed as
a multi-source generalization of the classical result
of Shannon~\cite{Shannon:IT}, 
which says that, loosely speaking, a discrete memoryless source
with known law can be losslessly reproduced if and only
if the data rate exceeds the entropy of the source. 
Shannon too studied a generalization of this
result, albeit in a different direction. 
He studied the problem of reproducing a source
imperfectly, subject to a minimum fidelity constraint, and
showed that the required rate is given by the well-known 
rate-distortion formula~\cite{Shannon:IT,Shannon:RD}. 
One of the central problems
of Shannon theory is to understand the limits of source
coding for models that combine the two generalizations.
That is, we seek to determine the rates required to
reproduce two correlated sources, each subject to a 
fidelity constraint, when the sources are encoded separately
(see Fig.~\ref{MTSC:basic:fig}). 
Determining the set
of achievable rates and distortions for this
setup is often
called the \emph{multiterminal source-coding problem}, 
even though this name suggests a more elaborate network
topology. This problem has been unsolved for some time.

\begin{figure}
\begin{center}
\scalebox{.9}{\input{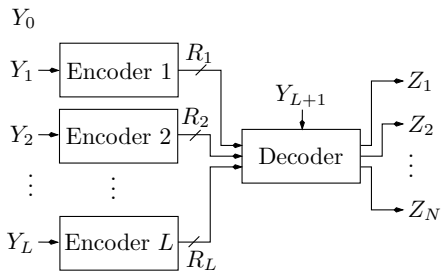}}
\end{center}
\caption{A general model.}
\label{MTSC:fig}
\end{figure}

The model we consider in this paper is slightly
more general and is depicted in Fig.~\ref{MTSC:fig}.
Beyond considering an arbitrary number of encoders, $L$,
we also allow for a hidden source, $Y_0$, which is
not directly observed by any encoder or the decoder, 
and a ``side information'' source, $Y_{L+1}$, which is observed by the
decoder but not by any encoder. We also permit arbitrary
functions of the sources to be reproduced, in addition to,
or in place of, the sources themselves.
We will therefore use $Z_1, Z_2$, etc., to denote
the instantaneous estimates instead of 
$\hat{Y}_1, \hat{Y}_2$, etc., as before.
In this paper, we will refer to this more general problem as
the multiterminal source-coding problem.

One might doubt the wisdom of embellishing the model
when even the basic form shown in Fig.~\ref{MTSC:basic:fig}
is unsolved. But one of the contributions of this paper is
to show that far from obscuring the problem,
the added generality actually illuminates it.
Of course, the more general problem is also unsolved.

Many special cases have been solved, however. For these,
the reader is referred to the classical papers of
Slepian and Wolf~\cite{Slepian:SW}, mentioned earlier;
Wyner~\cite{Wyner:SCSI};
Ahlswede and K\"{o}rner~\cite{Ahlswede:SCSI};
Wyner and Ziv~\cite{Wyner:WZ};
K\"{o}rner and Marton~\cite{Korner:ModTwo}; and
Gel`fand and Pinsker~\cite{Gelfand:CEO};
and to the more recent papers of
Berger and Yeung~\cite{Berger:BY};
Gastpar~\cite{Gastpar:WZ};
Oohama~\cite{Oohama:CEO:Region};
and Prabhakaran, Tse, and Ramchandran~\cite{Prabhakaran:ISIT04}.
While all of these papers contain conclusive results,
these results 
are established using coding theorems that are
tailored to the special cases under consideration.

The solutions to these solved special cases suggest a 
coding technique
for the general model~\cite{Berger:MTSC,Tung:PHD}. 
The idea is this. Each encoder
first quantizes its observation as in single-user
rate-distortion theory. The quantized processes
are then losslessly communicated to the decoder using
the binning scheme of Cover~\cite{Cover:Bin}.
The decoder uses the quantized processes to produce
the desired estimates. The set of rate-distortion
vectors that can be achieved using this scheme is
described in Section~\ref{relation}. This inner
bound to the rate-distortion region is tight in
all of the special cases listed above except
that of K\"{o}rner and Marton~\cite{Korner:ModTwo}.
Indeed, the K\"{o}rner-Marton problem seems to 
require a custom coding technique that relies on
the problem's unique structure. This suggests that
the multiterminal source-coding
problem may not have a classical single-letter
solution.

We attack this problem, therefore, by proving single-letter
inner and outer bounds on the rate-distortion
region. The best inner bound in the literature has
just been described.
The best outer bound, which is due to Berger~\cite{Berger:MTSC}
and Tung~\cite{Tung:PHD},
is described in Section~\ref{relation}.
In light of the result of K\"{o}rner and Marton, it is clear
that the two bounds must not coincide in all cases.
This gap cannot be entirely attributed to the
inner bound, however, as
there are instances of the problem that can be
solved from first principles for which the Berger-Tung outer bound
is strictly bigger than the true rate-distortion region (see
Section~\ref{contrived} of this paper).

Our aim is to provide an improved outer bound
for the problem. We prove such a bound 
in the next section,
following a precise formulation of the problem. We show that
our bound is contained in (i.e., subsumes)
the Berger-Tung outer bound in
Section~\ref{relation}.  In that section, we also provide several 
examples for which the containment is strict.

One example is the binary erasure version of the
``CEO problem,'' the general version of which was
introduced by Berger, Zhang, and
Viswanathan~\cite{Berger:CEO}. The CEO problem is a special
case of the multiterminal source-coding problem in which
the observed processes $Y_1,\ldots,Y_L$ are conditionally
independent given the hidden process $Y_0$ and in which
the decoder (the CEO) is only interested in estimating
the hidden process\footnote{This definition is not as
restrictive as it might seem. Indeed, any instance of
the multiterminal source-coding problem with a single
distortion constraint can be transformed into an instance
of the CEO problem without changing the rate-distortion
region by lumping $Y_1,\ldots,Y_L$ into $Y_0$ and redefining
the distortion measure as needed. Nonetheless, it defines
a useful special case.}. Berger, Zhang, 
and Viswanathan characterize the tradeoff
between sum rate and Hamming distortion in the 
high-rate and many-encoder
limit. Gel`fand and Pinsker~\cite{Gelfand:CEO}
had earlier found the 
rate region in the lossless reproduction case.
We consider the problem in which $Y_0$ is binary and uniform,
and the encoders observe $Y_0$ through independent
binary erasure channels. The decoder reproduces 
$Y_0$ subject to a constraint on the ``erasure
distortion'' (see Section~\ref{BECEO:subsection} or 
Cover and Thomas~\cite[p.~370]{Cover:IT}). For this
problem, we show that our outer bound is tight in the
sum rate for any number of users. In contrast,
the Berger-Tung outer bound contains points whose
sum rate is strictly smaller than the optimum.

In our view, this result is
of interest in its own right.
The binary erasure CEO problem arises naturally in sensor
networks in which the sensors occasionally ``sleep''
to conserve energy. This application is described
in Section~\ref{BECEO:subsection}. 
The result also provides an example for which
the binning-based coding scheme mentioned earlier
is optimum. Finally, this is one of relatively few
conclusive results for the multiterminal source-coding
problem in general, and the CEO problem in particular.
These problems are considered sufficiently difficult
that it is worth reporting solutions to special cases.

One of the few other conclusive results available is
for the Gaussian version of the CEO problem, which was
first studied by Viswanathan and Berger~\cite{Viswanathan:CEO}.
Here the encoders observe a hidden Gaussian source through
independent Gaussian additive-noise channels. The distortion
measure is expected squared error. The rate-distortion
region for this problem was recently found by 
Oohama~\cite{Oohama:Africa,Oohama:CEO:Region} and independently by 
Prabhakaran, Tse, and Ramchandran~\cite{Prabhakaran:ISIT04}.
We show that the converse result of these four authors
can be recovered from our single-letter outer bound, while the 
Berger-Tung outer bound contains points that lie outside the
true rate-distortion region.

The converse results used to solve all of the other special
cases mentioned so far are also consequences of our bound. This is 
discussed in Section~\ref{recover}. Our outer bound
therefore serves to unify most of what is known about
the nonexistence of multiuser source codes. 
This unification is noteworthy in the case of 
Oohama~\cite{Oohama:CEO:Region}
and Prabhakaran, Tse, and Ramchandran~\cite{Prabhakaran:ISIT04}
because the connection between their remarkable converse
result and the classical discrete results in this area
is not immediately apparent.
As we will see, subject to some technical caveats,
most of the key results in multiterminal
source coding can be recovered by combining the general inner
bound described earlier with the outer bound described next.

\section{Formulation and Main Result}

We work exclusively in discrete time.
We use uppercase letters to denote random variables and
vectors, lowercase letters to denote their realizations,
and script letters to denote their ranges. Let 
$\{Y^n_0(t), Y_1^n(t),\ldots,Y_L^n(t),Y_{L+1}^n(t)\}_{t = 1}^n$ be a 
vector-valued,
finite-alphabet memoryless source. For $A \subset \{1,\ldots,L\}$, we
denote $(Y_\ell^n(t))_{\{\ell \in A\}}$ by $\mathbf{Y}^n_A(t)$. If 
$A = \{1,\ldots,L\}$, we write this simply as $\mathbf{Y}^n(t)$.
In this context,
the set $A^c$ should be interpreted as 
$\{1,\ldots,L\}\backslash A$
rather than $\{0,\ldots,L+1\}\backslash A$.
When $A = \{\ell\}$, we shall write 
$Y_\ell^n(t)$ and $Y_{\ell^c}^n(t)$ in place of 
$Y_{\{\ell\}}(t)$ and $Y_{\{\ell\}^c}(t)$, respectively. Also,
we use $Y^n_\ell(t_1:t_2)$ to denote 
$\{Y^n_\ell(t)\}_{t = t_1}^{t_2}$,
$Y_\ell^n$ to denote $Y_\ell^n(1:n)$,
and $Y^n_\ell(t^c)$ to denote 
\begin{equation*}
(Y^n_\ell(1), \ldots, Y^n_\ell(t-1),Y^n_\ell(t+1), \ldots, Y^n_\ell(n)).
\end{equation*}
Similar notation will be used for other vectors that appear later.

\begin{figure}
\begin{center}
\makebox[3.3in][l]{\scalebox{.9}{\input{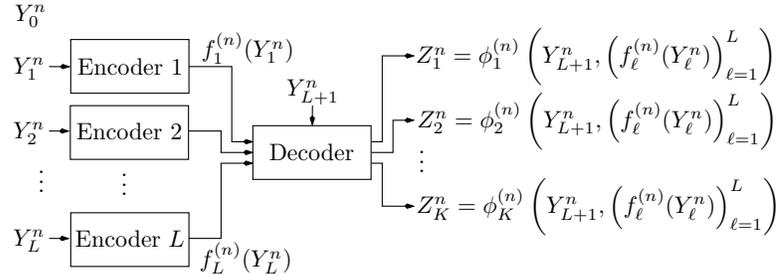}}}
\end{center}
\caption{Notation for the encoding and decoding rules.}
\label{rules}
\end{figure}

The notation for the 
encoding and decoding rules is shown in Fig.~\ref{rules}.
For each $\ell$ in $\{1,\ldots,L\}$, encoder $\ell$ observes 
$Y_\ell^n$,
then employs a mapping 
\begin{equation*}
f_\ell^{(n)} : \mathcal{Y}_\ell^n \rightarrow
\left\{1,\ldots,M_\ell^{(n)}\right\} 
\end{equation*}
to convey information about it to the decoder. 
The decoder observes $Y_{L+1}^n$ and uses it and the received messages to 
estimate $K$ functions of the vector-valued source according to the mappings
\begin{equation*}
\varphi_k^{(n)} : \mathcal{Y}_{L+1}^n \times 
  \prod_{\ell = 1}^L \left\{1,\ldots,M_\ell^{(n)}\right\} 
  \mapsto \mathcal{Z}_k^n \ \text{for $k = 1,\ldots,K$}.
\end{equation*}
We assume that $K$ distortion measures 
$d_k : \prod_{\ell = 0}^{L+1} \mathcal{Y}_\ell \times \mathcal{Z}_k \mapsto
\mathbb{R}_+$ are given. 

We mention at this point that while the generality of this
setup will be useful later when studying examples, it is not
needed to appreciate the bounding technique itself. The reader
is welcome to focus on the basic model shown in Fig.~\ref{MTSC:basic:fig}
for that purpose.
\begin{defn}
\label{achievable:defn}
The rate-distortion vector 
\begin{equation*}
(\mathbf{R},\mathbf{D}) = (R_1,R_2,\ldots,R_L,D_1,D_2,\ldots,D_K)
\end{equation*}
is \emph{achievable} if there exists a block length $n$, 
encoders $f_\ell^{(n)}$, and a decoder
\begin{equation*}
\left(\varphi_1^{(n)},\ldots,\varphi_K^{(n)}\right)
\end{equation*}
such that%
\footnote{All logarithms and exponentiations in this paper have
base $e$.}
\begin{equation}
\label{constraints}
\begin{split}
R_\ell & \ge \frac{1}{n} \log M_\ell^{(n)} \ \text{for all $\ell$, and} \\
D_k & \ge E\left[\frac{1}{n} \sum_{t = 1}^n d_k(Y^n_0(t),\mathbf{Y}^n(t),
Y_{L+1}^n(t), Z_k^n(t))\right] \ \text{for all $k$}.
\end{split}
\end{equation}
Let $\mathcal{RD}_\star$ be the
set of achievable rate-distortion vectors. Its
closure, $\overline{\mathcal{RD}_\star}$, is called the
\emph{rate-distortion region}.
\end{defn}
We will sometimes be concerned with projections of the
rate-distortion region. We denote these by, for example,
$\overline{\mathcal{RD}_\star} \cap \{R_1 = 0\}$, meaning
\begin{equation*}
\left\{(R_2,\ldots,R_L,D_1,\ldots,D_K) : (0,R_2,\ldots,R_L,D_1,\ldots,D_K)
  \in \overline{\mathcal{RD}_\star}\right\}.
\end{equation*}

In this paper, we view lossless compression as a limit of
lossy compression with the distortion tending to zero. More
precisely, if we wish to reproduce $Y_1$ losslessly, we will
set, say, $\mathcal{Z}_1 = \mathcal{Y}_1$ with $d_1$ equal
to Hamming distance, and then examine 
$\overline{\mathcal{RD}_\star} \cap \{D_1 = 0\}$.
This convention and Definition~\ref{achievable:defn} together
yield a notion of lossless compression that is weaker than the
one traditionally
used. It is common instead to require that for all sufficiently
large block lengths, there exists a code for which the probability
of correctly reproducing the entire vector $Y_1^n$ is arbitrarily
close to 1. But a weaker notion is desirable here since we are 
proving an outer bound or ``converse'' result.

To state our result,
let $Y_0,\ldots,Y_{L+1}$ be generic random variables with
the distribution of the source at a single time.
Let $\Gamma_o$ denote
the set of finite-alphabet random variables $\gamma = (U_1,\ldots,U_L,
Z_1,\ldots,Z_K,W,T)$ satisfying
\begin{enumerate}
\item[(\emph{i})]
$(W,T)$ is independent of $(Y_0,\mathbf{Y},Y_{L+1})$,
\item[(\emph{ii})]
$U_\ell \markov (Y_\ell,W,T) \markov (Y_0,\mathbf{Y}_{\ell^c},Y_{L+1},
  \mathbf{U}_{\ell^c}$), shorthand for
``$U_\ell$, $(Y_\ell,W,T)$ and 
$(Y_0,\mathbf{Y}_{\ell^c},Y_{L+1},\mathbf{U}_{\ell^c})$
form a Markov chain in this order'', for all $\ell$, and
\item[(\emph{iii})]
$(Y_0,\mathbf{Y},W) \markov (\mathbf{U},Y_{L+1},T) \markov \mathbf{Z}$.
\end{enumerate}
It is straightforward to verify that $\Gamma_o$ is
precisely the set of finite-alphabet random variables $(U_1,\ldots,U_L,
Z_1,\ldots,Z_K,W,T)$ whose joint distribution with
$(Y_0,\mathbf{Y},Y_{L+1})$ factors as
\begin{equation*}
p(y_0,\mathbf{y},y_{L+1},\mathbf{u},\mathbf{z},w,t) =
p(y_0,\mathbf{y},y_{L+1})p(w,t)\prod_{\ell = 1}^L p(u_\ell|y_\ell,w,t)
  p(\mathbf{z}|\mathbf{u},y_{L+1},t).
\end{equation*}
This description is helpful in that it suggests
a parametrization of the space $\Gamma_o$.

Let $\chi$ denote the set of finite-alphabet random variables $X$ with
the property that $Y_1,\ldots,Y_L$
are conditionally independent given $(X,Y_{L+1})$. Note 
that $\chi$ is nonempty since it contains, e.g., 
$X = (Y_1,\ldots,Y_{L-1})$.

There are many ways of coupling a given
$X$ in $\chi$ and $\gamma$ in $\Gamma_o$. 
In this paper, we shall only consider the unique coupling for which
$X \markov (Y_0,\mathbf{Y},Y_{L+1}) \markov \gamma$, which
we call the \emph{Markov coupling.} 
Whenever the joint distribution of
$X$, $(Y_0, \mathbf{Y}, Y_{L+1})$, and 
$\gamma$ arises, we assume that this coupling is in effect.

It is evident from the definition of $\chi$ that there is
considerable latitude in choosing how $X$ depends on
$Y_0$. This is because the sole constraint on the choice
of $X$ only depends on the joint distribution of $X$
and $(\mathbf{Y},Y_{L+1})$. But as the following 
definition makes clear, this freedom is inconsequential
since our outer bound only depends on the distributions
of $(Y_0,\mathbf{Y},Y_{L+1},\gamma)$ and $(X,\mathbf{Y},Y_{L+1},\gamma)$
separately.

\begin{defn}
\label{Me:defn}
Let
\begin{align*}
\mathcal{RD}_o(X,\gamma) & = \Bigg\{(\mathbf{R},\mathbf{D}) : 
\sum_{\ell \in A} R_\ell \ge I(X;\mathbf{U}_A|\mathbf{U}_{A^c},Y_{L+1},T) \\
& \phantom{= \Bigg\{} 
  \mbox{} + \sum_{\ell \in A} I(Y_\ell;U_\ell|X,Y_{L+1},W,T) 
    \ \text{for all $A \subset \{1,\ldots,L\}$}, \\
& \phantom{= \Bigg\{} \ \text{and} \ D_k 
  \ge E[d_k(Y_0,\mathbf{Y},Y_{L+1},Z_k)] \ \text{for all $k$} \Bigg\}.
\end{align*}
Then define 
\begin{equation*}
\mathcal{RD}_o = \bigcap_{X \in \chi} \bigcup_{\gamma \in \Gamma_o} 
          \mathcal{RD}_o(X,\gamma).  
\end{equation*}
\end{defn}
The first theorem is our main result. 
\begin{theorem} 
\label{main}
The rate-distortion region is contained in $\overline{\mathcal{RD}_o}$.
In fact, $\mathcal{RD}_\star \subset \mathcal{RD}_o.$
\end{theorem}

\emph{Proof.} It suffices to show the second statement.
Suppose $(\mathbf{R},\mathbf{D})$ is
achievable. Let $f_1^{(n)}, \ldots, f_L^{(n)}$ be encoders
and $(\varphi_1^{(n)},\ldots,\varphi_K^{(n)})$ a decoder 
satisfying~(\ref{constraints}).
Take any $X$ in $\chi$ and augment the sample space to
include $X^n$ so that 
\begin{equation*}
(X^n(t),Y_0^n(t),\mathbf{Y}^n(t),Y_{L+1}^n(t))
\end{equation*}
is independent over
$t$. Next let $T$ be uniformly distributed over $\{1,\ldots,n\}$,
independent of $X^n$, $Y_0^n$, $\mathbf{Y}^n$, and $Y_{L+1}^n$. 
Then define
\begin{align*}
X & = X^n(T) \\
Y_\ell & = Y_\ell^n(T) \ \text{for each $\ell$ in $\{0,\ldots,L+1\}$} \\
U_\ell & = \left(f_\ell^{(n)}(Y_\ell^n), X^n(1:T-1), 
     Y_{L+1}^n(T^c)\right) \ \text{for each $\ell$} \\
Z_k & = Z_k^n(T) \ \text{for each $k$} \\
W & = (X^n(T^c),Y^n_{L+1}(T^c)).
\end{align*}
It can be verified that $\gamma = (\mathbf{U},\mathbf{Z},W,T)$ 
is in $\Gamma_o$ and that,
together with $Y_0$, $\mathbf{Y}$, $Y_{L+1}$, and $X$, it satisfies the 
Markov coupling. It suffices to show that $(\mathbf{R},\mathbf{D})$
is in $\mathcal{RD}_o(X,\gamma)$.
First, note that~(\ref{constraints}) implies
\begin{equation*}
D_k \ge E[d_k(Y_0^n(T),\mathbf{Y}^n(T),Y_{L+1}^n(T),Z^n_k(T))] 
  \ \text{for all $k$}, 
\end{equation*}
i.e.,
\begin{equation*}
D_k \ge E[d_k(Y_0,\mathbf{Y},Y_{L+1},Z_k)] \ \text{for all $k$}.
\end{equation*}
Second, let $A \subset \{1,\ldots,L\}$. Then
by the cardinality bound on entropy,
\begin{equation*}
n \sum_{\ell \in A} R_\ell \ge 
   H\left(\left(f_\ell^{(n)}(Y_\ell^n)\right)_{\ell \in A}\right).
\end{equation*}
Since conditioning reduces entropy, this implies
\begin{align}
\nonumber
n \sum_{\ell \in A} R_\ell & \ge 
        H\left(\left(f_\ell^{(n)}(Y_\ell^n)\right)_{\ell \in A} 
   \Big|\left(f_\ell^{(n)}(Y_\ell^n)\right)_{\ell \in A^c},Y_{L+1}^n\right) \\
\label{blockchain}
    & = I\left(X^n, \mathbf{Y}_A^n; 
      \left(f_\ell^{(n)}(Y_\ell^n)\right)_{\ell \in A} \Big|
         \left(f_\ell^{(n)}(Y_\ell^n)\right)_{\ell \in A^c},Y_{L+1}^n\right).
\end{align}
By the chain rule for mutual information,
\begin{multline}
\label{twoterms}
 I\left(X^n, \mathbf{Y}_A^n ; 
    \left(f_\ell^{(n)}(Y_\ell^n)\right)_{\ell \in A} \Big|
        \left(f_\ell^{(n)}(Y_\ell^n)\right)_{\ell \in A^c},Y_{L+1}^n\right) \\
 = I\left(X^n; \left(f_\ell^{(n)}(Y_\ell^n)\right)_{\ell \in A} \Big|
         \left(f_\ell^{(n)}(Y_\ell^n)\right)_{\ell \in A^c},Y_{L+1}^n\right) \\
 + I\left(\mathbf{Y}_A^n; \left(f_\ell^{(n)}(Y_\ell^n)\right)_{\ell \in A} 
  \Big| \left(f_\ell^{(n)}(Y_\ell^n)\right)_{\ell \in A^c},X^n,Y_{L+1}^n\right).
\end{multline}
Applying the chain rule again gives
\begin{align*}
 & I\left(X^n; \left(f_\ell^{(n)}(Y_\ell^n)\right)_{\ell \in A}\Big| 
        \left(f_\ell^{(n)}(Y_\ell^n)\right)_{\ell \in A^c},Y_{L+1}^n\right) \\
 & = \sum_{t = 1}^n I\left(X^n(t); 
     \left(f_\ell^{(n)}(Y_\ell^n)\right)_{\ell \in A}\Big|
    \left(f_\ell^{(n)}(Y_\ell^n)\right)_{\ell \in A^c}, 
    X^n(1:t-1), Y_{L+1}^n\right).
\end{align*}
Consider next the second term on the right-hand side
of~(\ref{twoterms}). Since $X \in \chi$, 
\begin{multline*}
I\left(\mathbf{Y}_A^n; \left(f_\ell^{(n)}(Y_\ell^n)\right)_{\ell \in A}
  \Big| \left(f_\ell^{(n)}(Y_\ell^n)\right)_{\ell \in A^c},
         X^n,Y_{L+1}^n\right) \\
  = \sum_{\ell \in A} I\left(Y_\ell^n; f_\ell^{(n)}(Y_\ell^n)
       \Big|X^n,Y_{L+1}^n\right).
\end{multline*}
Applying the chain rule once more gives
\begin{multline*}
I\left(Y_\ell^n; f_\ell^{(n)}(Y_\ell^n)\Big|X^n,Y_{L+1}^n\right) = \\
\sum_{t = 1}^n I\left(Y_\ell^n(t); f_\ell^{(n)}(Y_\ell^n)\Big|X^n,
   Y_\ell^n(1:t-1), Y_{L+1}^n\right).
\end{multline*}
But 
\begin{multline*}
I\left(Y_\ell^n(t); f_\ell^{(n)}(Y_\ell^n)\Big|X^n,Y_\ell^n(1:t-1),
    Y_{L+1}^n\right) \\
  \mbox{} + I(Y_\ell^n(t);Y_\ell^n(1:t-1)|X^n,Y_{L+1}^n) \\
 =  I\left(Y_\ell^n(t);f_\ell^{(n)}(Y_\ell^n)\Big| X^n,Y_{L+1}^n\right) \\
  \mbox{} + I\left(Y_\ell^n(t);Y_\ell^n(1:t-1)\Big|f_\ell^{(n)}(Y_\ell^n),
   X^n,Y_{L+1}^n\right),
\end{multline*}
and the second term on the left-hand side is zero. Thus
\begin{multline*}
I\left(Y_\ell^n(t); f_\ell^{(n)}(Y_\ell^n)\Big|X^n,Y_\ell^n(1:t-1),
  Y_{L+1}^n\right) \ge \\
  I\left(Y_\ell^n(t);f_\ell^{(n)}(Y_\ell^n)\Big| X^n,Y_{L+1}^n\right).
\end{multline*}
Substituting the results of these various calculations into~(\ref{blockchain})
gives
\begin{equation}
\begin{split}
\label{substitute}
& \sum_{\ell \in A} R_\ell \ge \\
 & \frac{1}{n} \sum_{t = 1}^n \Bigg[I\left(X^n(t); 
           \left(f_\ell^{(n)}(Y_\ell^n)\right)_{\ell \in A}\Big|
  \left(f_\ell^{(n)}(Y_\ell^n)\right)_{\ell \in A^c}, 
         X^n(1:t-1),Y_{L+1}^n\right) \\
 & \phantom{\ge} + \sum_{\ell \in A}
      I\left(Y_\ell^n(t);f_\ell^{(n)}(Y^n_\ell)\Big|
       X^n(t),X^n(t^c),Y_{L+1}^n(t), Y_{L+1}^n(t^c)\right)\Bigg].
\end{split}
\end{equation}
If $A^c$ is nonempty, this can be rewritten as
\begin{align*}
\sum_{\ell \in A} R_\ell
& \ge I(X^n(T);\mathbf{U}_A|\mathbf{U}_{A^c},Y_{L+1}^n(T),T) \\
& \phantom{=} + \sum_{\ell \in A}
  I(Y_\ell^n(T);U_\ell|X^n(T),X^n(T^c),Y_{L+1}^n(T),Y_{L+1}^n(T^c),T) \\
& = I(X;\mathbf{U}_A|\mathbf{U}_{A^c},Y_{L+1},T) + \sum_{\ell \in A}
    I(Y_\ell;U_\ell|X,Y_{L+1},W,T).
\end{align*}
The case $A = \{1,\ldots,L\}$ is handled separately. In this case,
observe that 
\begin{align*}
 & I\left(X^n(t); \left(f_\ell^{(n)}(Y_\ell^n)\right)_{\ell \in A} 
     \Big| \left(f_\ell^{(n)}(Y_\ell^n)\right)_{\ell \in A^c}, X^n(1:t-1),
       Y_{L+1}^n\right) \\
 & = I\left(X^n(t); \left(f_\ell^{(n)}(Y_\ell^n)\right)_{\ell \in A}
      \Big|  X^n(1:t-1), Y_{L+1}^n\right)  \\
 & = I\left(X^n(t); \left(f_\ell^{(n)}(Y_\ell^n)\right)_{\ell \in A}
      \Big|  X^n(1:t-1), Y_{L+1}^n\right) \\
 & \phantom{= I\Bigg(} + I(X^n(t);X^n(1:t-1),Y_{L+1}^n(t^c)|Y_{L+1}^n(t)) \\
 & = I\left(X^n(t);\left(f_\ell^{(n)}(Y_\ell^n)\right)_{\ell \in A},
     X^n(1:t-1),Y_{L+1}^n(t^c)\Big| Y_{L+1}^n(t)\right).
\end{align*}
Substituting this into~(\ref{substitute}) and proceeding as in the
$A^c \ne \emptyset$ case completes the proof. \QED

It is worth noting that the proof uses classical
techniques. Most of the manipulations in the latter part of the
proof can be viewed as versions
of the chain rule
for mutual information. Since this chain rule holds in
abstract spaces~\cite[(3.6.6)]{Pinsker:Stability}, the proof
can be readily extended to more general alphabets.

The key step in the proof is the introduction of
$X^n$ in~(\ref{blockchain}). Unlike the other
auxiliary random variables, $X^n$ does not 
represent a component of the code. Rather, it is used to 
aid the analysis by inducing conditional independence among 
the messages sent by the encoders. This technique of augmenting
the source to induce conditional independence was pioneered
by Ozarow~\cite{Ozarow:MD}, who used it to solve the Gaussian 
two-descriptions problem.
Wang and Viswanath~\cite{Wang:MD} used it to determine the
sum rate of the Gaussian vector multiple-descriptions problem with 
individual and central decoders. It was also used by Wagner,
Tavildar, and Viswanath~\cite{Wagner:Gaussian:Two} to solve the Gaussian
two-terminal source-coding problem. A step that is similar 
to~(\ref{blockchain}) appeared in 
Gel`fand and Pinsker~\cite{Gelfand:CEO} and 
in later papers on the Gaussian CEO
problem~\cite{Oohama:CEO:Region,Prabhakaran:ISIT04},
although in these works $X^n$ is part of the source,
so no augmentation is involved.

The significance of conditional independence has long
been known in the related field of distributed detection
(e.g.,~\cite{Tsitsiklis:Complexity}). Given the
similarity between distributed detection and the
multiterminal source-coding problem, one expects
conditional independence to play a significant
role here as well. Indeed,
most conclusive results for the 
multiterminal source-coding problem require
a conditional independence 
assumption~\cite{Gelfand:CEO,Berger:CEO,Gastpar:WZ,
Oohama:CEO:Region,Prabhakaran:ISIT04}. The motivation
for introducing $X^n$ is that it allows one to apply
the approach used in these works to problems that
lack conditional independence.

We do not consider the problem of computing $\mathcal{RD}_o$ in this
paper. Note that we have not specified the
alphabet sizes of the auxiliary random variables $\mathbf{U}$,
$W$, and $T$. As such, the outer bound provided by Theorem~\ref{main}
is not computable~\cite[p.~259]{CK:IT}
in the present form. One might question 
the utility of an
outer bound that cannot be computed. The remainder of the paper,
however, will show that the bound is
still useful as a theoretical tool. In addition, cardinality bounds
might be found later, although obtaining such bounds appears to
be more difficult in this case than for related bounds.

It should be mentioned that the time-sharing variable
$T$ is unnecessary; it
can be absorbed into the other variables. We have included it to
ease the comparison with existing inner and outer bounds, to which
we turn next.

\section{Relation to Existing Bounds}
\label{relation}

The coding scheme described in the introduction
gives rise to the following inner bound on the
rate-distortion region.
\begin{defn}
\label{BT:inner:defn}
Let $\Gamma_i^{BT}$ denote the set of finite-alphabet random
variables 
\begin{equation*}
\gamma = (U_1,\ldots,U_L, Z_1,\ldots,Z_K,T)
\end{equation*}
satisfying
\begin{enumerate}
\item[(\emph{i})] $T$ is independent of $(Y_0,\mathbf{Y},Y_{L+1})$,
\item[(\emph{ii})] $U_\ell 
        \markov (Y_\ell,T) \markov (Y_0,\mathbf{Y}_{\ell^c},Y_{L+1},
       \mathbf{U}_{\ell^c})$ for all $\ell$, and
\item[(\emph{iii})] $(Y_0,\mathbf{Y}) \markov (\mathbf{U},Y_{L+1},T) 
             \markov \mathbf{Z}$.
\end{enumerate}
Then define
\begin{align*}
\mathcal{RD}_i^{BT}(\gamma) & = \Bigg\{ (\mathbf{R},\mathbf{D}) : 
\sum_{\ell \in A} R_\ell \ge I(\mathbf{Y}_A;\mathbf{U}_A|%
       \mathbf{U}_{A^c},Y_{L+1},T) \ \text{for all $A$}, \\
  & \phantom{= \Bigg\{} \ \text{and} \ 
 D_k \ge E[d_k(Y_0,\mathbf{Y},Y_{L+1},Z_k)] \ \text{for all $k$} \Bigg\}.
\end{align*}
Finally, let
\begin{equation*}
\mathcal{RD}_i^{BT} = \bigcup_{\gamma \in \Gamma_i^{BT}}
\mathcal{RD}_i^{BT}(\gamma).
\end{equation*}
\end{defn}
\begin{prop}[\cite{Berger:MTSC,Tung:PHD}]
\label{BT:inner}
$\overline{\mathcal{RD}_i^{BT}} \subset \overline{\mathcal{RD}_\star}$.
\end{prop}
In Appendix~\ref{closed} we show that $\mathcal{RD}_i^{BT}$
is in fact closed. We call $\overline{\mathcal{RD}_i^{BT}}$ the Berger-Tung
\cite{Berger:MTSC,Tung:PHD} 
inner bound, since although these authors prove a bound that is less general
than the one given here, their proof can be extended to prove
Proposition~\ref{BT:inner}.
See Chen \emph{et al.}~\cite{Chen:CEO:Bounds} or 
Gastpar~\cite{Gastpar:WZ} for recent sketches of the
proof that accommodate some of the generalizations
included here.

To understand the difference between $\mathcal{RD}_i^{BT}$ and
$\mathcal{RD}_o$, suppose that 
\begin{equation*}
(\mathbf{U},\mathbf{Z},W,T)
\end{equation*}
is in $\Gamma_o$ and $W$ is deterministic. Then
$(\mathbf{U},\mathbf{Z},T)$ is in $\Gamma_i^{BT}$, and for
all $A \subset \{1,\ldots,L\}$ and all $X \in \chi$, 
\begin{align*}
& I(X;\mathbf{U}_A|\mathbf{U}_{A^c},Y_{L+1},T) + 
    \sum_{\ell \in A} I(Y_\ell;U_\ell|X,Y_{L+1},W,T) \\
 & = I(X;\mathbf{U}_A|\mathbf{U}_{A^c},Y_{L+1},T) + 
     I(\mathbf{Y}_A;\mathbf{U}_A|\mathbf{U}_{A^c},X,Y_{L+1},W,T) \\
 & = I(X;\mathbf{U}_A|\mathbf{U}_{A^c},Y_{L+1},T) + 
     I(\mathbf{Y}_A;\mathbf{U}_A|\mathbf{U}_{A^c},X,Y_{L+1},T) \\
 & = I(X,\mathbf{Y}_A;\mathbf{U}_A|\mathbf{U}_{A^c},Y_{L+1},T) \\
 & = I(\mathbf{Y}_A;\mathbf{U}_A|\mathbf{U}_{A^c},Y_{L+1},T).
\end{align*}
Thus 
\begin{equation}
\label{tworegions}
\mathcal{RD}_o(X,\mathbf{U},\mathbf{Z},W,T) =
\mathcal{RD}_i^{BT}(\mathbf{U},\mathbf{Z},T).
\end{equation}
Conversely, if $(\mathbf{U},\mathbf{Z},T)$ is in
$\Gamma_i^{BT}$, then for any deterministic $W$,
$(\mathbf{U},\mathbf{Z},W,T)$ is in $\Gamma_o$
and~(\ref{tworegions}) holds for any $X$.
It follows that $\mathcal{RD}_i^{BT}$ is equal to
$\mathcal{RD}_o$
with $W$ restricted to be deterministic in
the definition of $\Gamma_o$.

In particular, to obtain coincident inner
and outer bounds, it suffices to show that restricting $W$ to be
deterministic in the definition of $\Gamma_o$ does not reduce
$\mathcal{RD}_o$. We will see later how this can be accomplished
in several examples. Of course, it is not possible for the problem
solved by K\"{o}rner and Marton~\cite{Korner:ModTwo}, since they show
that the inner bound is not tight in that case.

The best outer bound in the literature
is the following.
\begin{defn}
\label{BT:outer:defn}
Let $\Gamma_o^{BT}$ denote the set of finite-alphabet random
variables $\gamma = (\mathbf{U},\mathbf{Z},T)$ satisfying
\begin{enumerate}
\item[(\emph{i})] $T$ is independent of $(Y_0,\mathbf{Y},Y_{L+1})$,
\item[(\emph{ii})] $U_\ell \markov (Y_\ell,T) 
     \markov (Y_0,\mathbf{Y}_{\ell^c},Y_{L+1})$ for all $\ell$, and
\item[(\emph{iii})] $(Y_0,\mathbf{Y}) \markov (\mathbf{U}, Y_{L+1},T) 
      \markov \mathbf{Z}$.
\end{enumerate}
Then let
\begin{align*}
\mathcal{RD}_o^{BT}(\gamma) & = \Bigg\{(\mathbf{R},\mathbf{D}) : 
\sum_{\ell \in A} R_\ell \ge I(\mathbf{Y};\mathbf{U}_A|
         \mathbf{U}_{A^c},Y_{L+1},T) \ 
  \text{for all $A$,} \\
 & \phantom{= \Bigg\{} 
   \ \text{and} \ D_k \ge E[d_k(Y_0,\mathbf{Y},Y_{L+1},Z_k)] \ 
  \text{for all $k$} \Bigg\}.
\end{align*}
Finally, let
\begin{equation*}
\mathcal{RD}_o^{BT} = \bigcup_{\gamma \in \Gamma_o^{BT}}
\mathcal{RD}_o^{BT}(\gamma).
\end{equation*}
\end{defn}
\begin{prop}[\cite{Berger:MTSC,Tung:PHD}]
$\mathcal{RD}_\star \subset \mathcal{RD}_{o}^{BT}$.
\end{prop}
As with the inner bound, Berger~\cite{Berger:MTSC} and
Tung~\cite{Tung:PHD} prove the result for a model
that is more restrictive than the one considered here,
but their proof can be extended
to this setup (c.f.~\cite{Chen:CEO:Bounds,Gastpar:WZ}).
The difference between $\mathcal{RD}_i^{BT}$ and
$\mathcal{RD}_o^{BT}$ is that
condition (\emph{ii}) has been weakened in the latter.
We next show that the Berger-Tung outer bound is subsumed by the 
one in the previous section.
\begin{prop}
\label{contain}
$\mathcal{RD}_o \subset \mathcal{RD}_o^{BT}$.
\end{prop}

\emph{Proof.}
First observe that for any $(\mathbf{U},\mathbf{Z},W,T)$ in
$\Gamma_o$,
\begin{equation*}
U_\ell \markov (Y_\ell,W,T) \markov (Y_0,\mathbf{Y}_{\ell^c},Y_{L+1}) 
   \ \text{for each $\ell$}.
\end{equation*}
Since
$(Y_0,\mathbf{Y}_{\ell^c},Y_{L+1}) \markov (Y_\ell,T) \markov W$, it holds
\begin{equation*}
U_\ell \markov (Y_\ell,T) \markov (Y_0,\mathbf{Y}_{\ell^c},Y_{L+1})
  \ \text{for each $\ell$}.
\end{equation*}
Thus $(\mathbf{U},\mathbf{Z},T)$ is in $\Gamma_o^{BT}$ and in
particular,
\begin{equation*}
\mathcal{RD}_o(\mathbf{Y},\mathbf{U},\mathbf{Z},W,T) = 
  \mathcal{RD}_o^{BT}(\mathbf{U},\mathbf{Z},T).
\end{equation*}
It follows that
\begin{equation*}
\mathcal{RD}_o \subset \bigcup_{\gamma \in \Gamma_o} \mathcal{RD}_o(\mathbf{Y},
   \gamma) \subset \bigcup_{\gamma \in \Gamma_o^{BT}} 
    \mathcal{RD}_o^{BT}(\gamma) = \mathcal{RD}_o^{BT}.
\end{equation*}
\QED 

The proof reveals that 
$\mathcal{RD}_o$ improves upon $\mathcal{RD}_o^{BT}$ in
two ways. The first is that $\mathcal{RD}_o$ allows
for optimization over $X$ while $\mathcal{RD}_o^{BT}$
effectively requires the choice $X = \mathbf{Y}$. The
second is that $\Gamma_o$ is ``smaller'' than 
$\Gamma_o^{BT}$ in the sense that if 
$(\mathbf{U},\mathbf{Z},W,T)$ is in $\Gamma_o$ then
$(\mathbf{U},\mathbf{Z},T)$ is in $\Gamma_o^{BT}$.
The balance of this section is devoted to showing
that these improvements make the containment
in Proposition~\ref{contain} strict in some
cases. As the reader will see, the former 
difference is entirely responsible for the gap 
that we expose between the two
bounds in our examples. We hasten to add, however, 
that the latter improvement is not an empty one in 
that Anantharam and Borkar~\cite{Anantharam:Mixtures}
have shown that there can exist a
$(\mathbf{U},\mathbf{Z},T)$ in $\Gamma_o^{BT}$
with the property that there does not exist a $W$ 
such that $(\mathbf{U},\mathbf{Z},W,T)$ is in $\Gamma_o$.
It is interesting to note that the 
Anantharam-Borkar example arose independently
of this work in the context of distributed
stochastic control.

We will exhibit three examples for which $\mathcal{RD}_o^{BT}$
strictly contains $\mathcal{RD}_o$.
The first is rather contrived and can be
solved from first principles. It is included to 
illustrate the difference between the two bounds.
\subsection{Toy Example}
\label{contrived}
Let $Y_{11}$, $Y_{12}$, $Y_{21}$, and $Y_{22}$ be independent
and identically distributed (i.i.d.) random variables,
uniformly distributed over $\{0,1\}$. Consider two encoders ($L = 2$) with
$Y_1 = (Y_{11},Y_{12})$ and $Y_2 = (Y_{21},Y_{22})$
(there is no hidden source or side information in this example).
We have a single distortion constraint ($K=1$)
with $\mathcal{Z}_1 = \{0,1\}^2$ and
\begin{equation*}
d_1(\mathbf{Y},Z_1) = 
\begin{cases}
0 & \text{if $Z_1 = (Y_{11},Y_{21})$ or $Z_1 = (Y_{12},Y_{22})$} \\
1 & \text{otherwise}.
\end{cases}
\end{equation*}
In words, the decoder attempts to guess either the first or the
second coordinate of both encoders' observations.
It incurs a distortion of zero if it guesses correctly
the same coordinate of the two sources and one otherwise.
Note that the decoder need not declare which coordinate
it is attempting to guess.
\begin{prop}
For this problem,
\begin{equation*}
 \overline{\mathcal{RD}_o} \cap \{D_1 = 0\} 
  \subset \{(R_1,R_2) : R_1 \ge \log 2, R_2 \ge \log 2 \}.
\end{equation*}
\end{prop}
\emph{Proof.} 
Suppose $(R_1,R_2,\epsilon)$ is in $\mathcal{RD}_o$, and
$\epsilon \le 1/2$. Observe that since $Y_1$ and $Y_2$
are independent, deterministic random variables are in
$\chi$. Thus there exists $\gamma$ in $\Gamma_o$ such that
\begin{align*}
\epsilon & \ge E[d_1(\mathbf{Y},Z_1)] \\
R_1 & \ge I(Y_1;U_1|W,T) \\
R_2 & \ge I(Y_2;U_2|W,T).
\end{align*}
By condition (\emph{ii}) defining $\Gamma_o$,
\begin{equation}
\label{repeatii}
U_1 \markov (Y_1,W,T) \markov (Y_2,U_2).
\end{equation}
Since $Y_2$ is independent of $(Y_1,W,T)$ in this
example, $Y_2$ must be
independent of $(Y_1,U_1,W,T)$. Thus
\begin{equation}
\label{toycondition}
I(Y_1;U_1|W,T) = I(Y_1;U_1|W,T,Y_{21} \ne Y_{22}).
\end{equation}
Likewise, $Y_1$ is independent of $(Y_2,U_2,W,T)$ and
hence given $(W,T)$, $Y_1$ is independent of $(Y_2,U_2)$.
This observation combined with~(\ref{repeatii}) implies
$(Y_1,U_1) \markov (W,T) \markov (Y_2,U_2)$. In particular,
$Y_1 \markov (U_1,W,T) \markov (Y_2,U_2)$. By condition
(\emph{iii}) defining $\Gamma_o$, $Y_1 \markov (Y_2,U_1,U_2,W,T) \markov
 (Y_2,Z_1)$. These last two chains imply that
\begin{equation*}
Y_1 \markov (U_1,W,T) \markov (Y_2,Z_1).
\end{equation*}
Thus conditioned on $(W,T)$ and the event 
$\{Y_{21} \ne Y_{22}\}$, we have $Y_1 \markov U_1 \markov (Y_2,Z_1)$. 
It follows that
\begin{align*}
& I(Y_1;U_1|W,T,Y_{21} \ne Y_{22}) \\
 & \ge I(Y_1;Y_2,Z_1|W,T,Y_{21} \ne Y_{22}) \\
  & = I(Y_1;Y_2,Z_1,d_1(\mathbf{Y},Z_1)|W,T,Y_{21} \ne Y_{22}) \\ 
  & \phantom{= I(}
   - I(Y_1;d_1(\mathbf{Y},Z_1)|W,T,Y_2,Z_1,Y_{21} \ne Y_{22}) \\
  & \ge H(Y_1|W,T,Y_{21} \ne Y_{22}) - H(Y_1|Y_2,Z_1,d_1(\mathbf{Y},Z_1),
     W,T,Y_{21} \ne Y_{22}) \\ 
  & \phantom{= I(}
   - H(d_1(\mathbf{Y},Z_1)|W,T,Y_2,Z_1,Y_{21} \ne Y_{22}), 
\end{align*}
since $d_1(\mathbf{Y},Z_1)$ is a function of $\mathbf{Y}$ and $Z_1$. Next,
observe that on the events $\{Y_{21} \ne Y_{22}\}$ and 
$\{d_1(\mathbf{Y},Z_1) = 0\}$,
$Y_2$ and $Z_1$ together must reveal one of the two bits of $Y_1$.
Thus
\begin{equation*}
H(Y_1|Y_2,Z_1,d_1(\mathbf{Y},Z_1) = 0,W,T,Y_{21} \ne Y_{22}) \le \log 2.
\end{equation*}
Continuing our chain of inequalities,
\begin{align}
\nonumber
 & I(Y_1;U_1|W,T,Y_{21} \ne Y_{22}) \\
\nonumber
  & \ge 2\log 2 - \log 2 \cdot
       \Pr(d_1(\mathbf{Y},Z_1) = 0|Y_{21} \ne Y_{22}) \\
\nonumber
   & \phantom{= I(}
    - (2 \log 2) \cdot 
      \Pr(d_1(\mathbf{Y},Z_1) = 1|Y_{21} \ne Y_{22}) \\
\nonumber
   &  \phantom{= I(}
   - H(d_1(\mathbf{Y},Z_1)|Y_{21} \ne Y_{22}) \\
\label{toyformula}
  & \ge \log 2 - (2\log 2) \cdot 
     \Pr(d_1(\mathbf{Y},Z_1) = 1|Y_{21} \ne Y_{22}) \\
\nonumber
   &  \phantom{= I(}
     - H(d_1(\mathbf{Y},Z_1)|Y_{21} \ne Y_{22}).
\end{align}
Now
\begin{multline*}
\frac{1}{2} H(d_1(\mathbf{Y},Z_1)|Y_{21} \ne Y_{22}) +
\frac{1}{2} H(d_1(\mathbf{Y},Z_1)|Y_{21} = Y_{22}) \\
   = H(d_1(\mathbf{Y},Z_1)|1(Y_{21} = Y_{22})) 
 \le H(d_1(\mathbf{Y},Z_1)) \le h(\epsilon),
\end{multline*}
where, here and throughout, $h(\cdot)$ is the binary entropy
function with natural logarithms. We conclude that
\begin{equation*}
H(d_1(\mathbf{Y},Z_1)|Y_{21} \ne Y_{22}) \le 2h(\epsilon).
\end{equation*}
Similarly,
\begin{equation*}
\Pr(d_1(\mathbf{Y},Z_1) = 1|Y_{21} \ne Y_{22}) \le 2\epsilon.
\end{equation*}
Substituting these two observations into~(\ref{toyformula}) and
recalling~(\ref{toycondition}) yields
\begin{equation*}
I(Y_1;U_1|W,T) \ge \log 2 - 4\epsilon \log 2 - 2h(\epsilon).
\end{equation*}
By symmetry, $I(Y_2;U_2|W,T)$ must satisfy the same
inequality. This implies the desired conclusion. \QED

It is easy to see that the point $(\log 2, \log 2, 0)$ is achievable.
Using rate $\log 2$,
each encoder can send, say, the first coordinate of
its observation. The decoder can then 
realize zero distortion by repeating the two bits it receives.
This fact and the above proposition together imply
\begin{align*}
\overline{\mathcal{RD}_o} \cap \{D_1 = 0\} = 
\overline{\mathcal{RD}_\star} \cap \{D_1 = 0\} = 
\{(R_1,R_2): R_1  \ge \log 2, R_2 \ge \log 2\}.
\end{align*}
In particular, $\mathcal{RD}_o$ is tight in the zero-distortion limit.
In contrast, we show next that the Berger-Tung outer bound is not.
\begin{prop}
The point $((3/4)\log 2, (3/4)\log 2, 0)$ is contained in $\mathcal{RD}_o^{BT}$.
\end{prop}
\emph{Proof.}
Let the random variable $W$ be uniformly distributed
over $\{1,2\}$, and let $U_1 = Y_{1W}$ and
$U_2 = Y_{2W}$. Let $Z_1 = (U_1,U_2)$. It is straightforward
to verify that $(U_1,U_2,Z_1)$ is in $\Gamma_o^{BT}$ (the
time-sharing random variable $T$ is unneeded 
and can be taken to be constant). Next note that
$E[d_1(\mathbf{Y},Z_1)] = 0$. Finally, one can compute
\begin{align*}
I(\mathbf{Y};U_1,U_2) & = \frac{5}{4}\log 2 \\
\intertext{and}
I(\mathbf{Y};U_1) & = \frac{1}{2}\log 2.
\end{align*}
This implies that
\begin{equation*}
I(\mathbf{Y};U_1|U_2) = I(\mathbf{Y};U_2|U_1) = \frac{3}{4} \log 2
 > \frac{5}{8}\log 2.
\end{equation*}
The conclusion follows. \QED

\subsection{Binary Erasure CEO Problem}
\label{BECEO:subsection}

Here $Y_0$ is uniformly distributed
over $\{-1,1\}$, and $Y_\ell = N_\ell \cdot Y_0$ for $\ell$ in
$\{1,\ldots,L\}$, where
$N_1,\ldots,N_L$ are i.i.d.\ with 
$0 < \Pr(N_1 = 0) = p < 1$ and 
$\Pr(N_1 = 1) = 1 - p$.
Let $\mathcal{Z}_1 = \{-1,0,1\}$. We will assume that
there is no side information and that
the decoder is only interested in reproducing the
hidden process
$Y_0$. We measure the fidelity of its reproduction
using a family of distortion
measures, $\{d^\lambda_1\}_{\lambda > 0}$, where
\begin{equation*}
d^\lambda_1(Y_0,\mathbf{Y},Z_1) = 
\begin{cases}
0 & \text{if $Y_0 = Z_1$} \\
1 & \text{if $Z_1 = 0$} \\
\lambda & \text{otherwise}.
\end{cases}
\end{equation*}
We are particularly interested in the large-$\lambda$ limit. In
this regime, $d^\lambda_1$ approximates the ``erasure distortion
measure''~\cite[p.~370]{Cover:IT},
\begin{equation*}
d^\infty_1(Y_0,\mathbf{Y},Z_1) =
\begin{cases}
0 & \text{if $Y_0 = Z_1$} \\
1 & \text{if $Z_1 = 0$} \\
\infty & \text{otherwise}.
\end{cases}
\end{equation*}
We use a finite approximation because an infinite distortion
measure causes difficulties in the proof of the Berger-Tung
inner bound.

This example is motivated by the following problem arising in
energy-limited sensor networks.
We seek to monitor a remote source, $Y_0$.
To this end, we deploy an array of sensors, each of which is
capable of observing the source with negligible probability
of error. To lengthen the lifetime of the network, each sensor
spends a fraction $p$ of the time in a low-power ``sleep''
state. We assume that the sensors cycle between
the awake and sleep states independently of each other and on a faster
time scale than the sampling; at each
discrete time, each sensor sleeps with probability $p$,
independently of the other sensors and the past.
Sensors do not make any observations while
they are asleep, resulting in erasures. We permit the coding
process to introduce additional erasures, but not
errors, yielding the erasure distortion measure.
What sum rate is required in order for the decoder to 
reproduce a fraction $1 - D$ of the $Y_0^n$ variables while
almost never making an error?
Of course, $D$ must satisfy $D \ge p^L$.

Define
\begin{equation*}
\mathcal{R}_\star(D,\lambda) = \inf\left\{ \sum_{\ell = 1}^L R_\ell :
 (R_1,\ldots,R_L,D) \in \overline{\mathcal{RD}_\star}(\lambda)\right\},
\end{equation*}
where $\overline{\mathcal{RD}_\star}(\lambda)$ is the rate-distortion
region when the distortion measure is $d_1^\lambda$. We define
$\mathcal{R}_o(D,\lambda)$ and $\mathcal{R}_i^{BT}(D,\lambda)$
analogously.

\begin{figure}
\begin{center}
\scalebox{.5}{\input{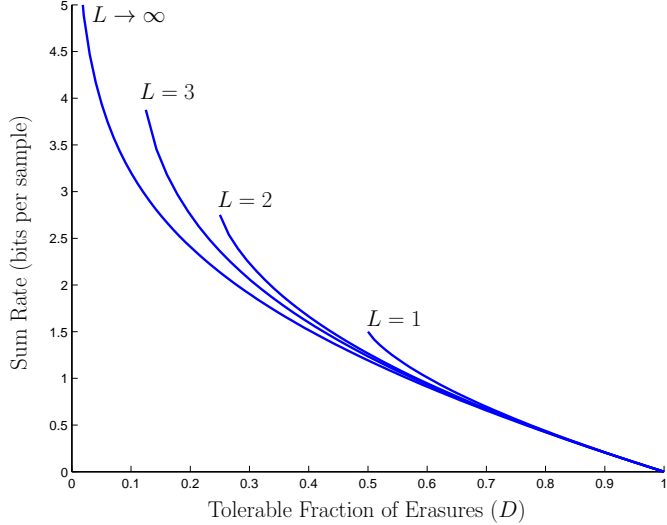}}
\end{center}
\caption{Sum rate for the binary erasure CEO problem with $p = 1/2$.}
\label{BECEO:figure}
\end{figure}

In Appendix~\ref{BECEO:achieve}, we show that if $p^L \le D \le 1$, then
\begin{equation}
\label{BECEO:sumrate}
\lim_{\lambda \rightarrow \infty} \mathcal{R}_i^{BT}(D,\lambda) \le
     (1-D)\log 2
 + L\left[h\left(D^{1/L}\right) -
(1-p)h\left(\frac{D^{1/L} - p}{1-p}\right)\right].
\end{equation}
In Appendix~\ref{BECEO:converse}, we show that the quantity
on the right-hand side is also a lower bound to
$\lim_{\lambda \rightarrow \infty} \mathcal{R}_o(D,\lambda)$.
Hence it must equal
$\lim_{\lambda \rightarrow \infty} \mathcal{R}_\star(D,\lambda)$.
That is, the improved outer bound and the Berger-Tung inner bound
together
yield a conclusive result for the sum rate of the 
binary erasure CEO problem. Evidently this problem was previously 
unsolved. In Appendix~\ref{BT:loose:BE}, we show that
$\overline{\mathcal{RD}_o^{BT}}$ contains points with a strictly 
smaller sum rate in general. Fig.~\ref{BECEO:figure} shows
the correct sum rate for $p = 0.5$ and several values of $L$.

\subsection{Gaussian CEO Problem~\cite{Viswanathan:CEO,Oohama:CEO,
Oohama:CEO:Region,Prabhakaran:ISIT04}}
\label{CEO:Gaussian}

We turn to a continuous example. Here
$Y_0,\ldots,Y_L$ are jointly Gaussian and
$Y_1,\ldots,Y_L$ are conditionally independent given $Y_0$. 
For $\ell \ge 1$, let us write $Y_\ell = Y_0 + N_\ell$, where
$Y_0,N_1,\ldots,N_L$ are mutually independent and
\begin{equation*}
E\left[N_\ell^2\right] = \sigma^2_\ell > 0 \ \text{for all $\ell$}.
\end{equation*}
We will denote the variance of $Y_0$ by $\sigma^2$.
Again there is no side information, and the decoder is only
interested in reproducing the hidden process $Y_0$,
\begin{equation*}
d_1(Y_0,\mathbf{Y},Z_1) = (Y_0 - Z_1)^2.
\end{equation*}
The rate-distortion region for this problem was
recently found by Oohama~\cite{Oohama:Africa,Oohama:CEO:Region}
and Prabhakaran, Tse, and 
Ramchandran~\cite{Prabhakaran:ISIT04}.
The two proofs are nearly the same, and build on 
earlier work of Oohama~\cite{Oohama:CEO}.
The primary contribution is the converse result,
which makes heavy use of the entropy power 
inequality~\cite[Theorem~16.6.3]{Cover:IT}.
The Berger-Tung inner bound is used for achievability.

It is straightforward to extend
Theorem~\ref{main} to this continuous
setting. A statement of the continuous
version is given in Appendix~\ref{Gaussian:proof},
where we also use the techniques of Oohama~\cite{Oohama:CEO:Region}
and Prabhakaran, Tse, and
Ramchandran~\cite{Prabhakaran:ISIT04} to prove the following.
\begin{prop} 
\label{Gaussian:Prop}
For the Gaussian CEO problem,
\begin{multline}
\label{Gaussian:result}
\mathcal{RD}_o \subset
  \Bigg\{(R_1,\ldots,R_L,D) \in \mathbb{R}^{L+1}_+ : \ \text{there exists} \
  (r_1,\ldots,r_L) \in \mathbb{R}^L_+ : \text{for all $A$},  \\
 \sum_{\ell \in A} R_\ell \ge \frac{1}{2} \log^+ \Bigg[\frac{1}{D}
 \Bigg( \frac{1}{\sigma^2} + \sum_{\ell \in A^c} 
   \frac{1 - \exp(-2r_\ell)}{\sigma^2_\ell}\Bigg)^{-1}\Bigg]
   + \sum_{\ell \in A} r_\ell \Bigg\},
\end{multline}
\end{prop}
where $\log^+ x = \max(\log x,0)$.
Since this expression equals
$\overline{\mathcal{RD}_\star}$~\cite{Oohama:CEO:Region}, we conclude
that $\mathcal{RD}_o$ is tight in this example. It also
follows that the converse result of Oohama~\cite{Oohama:CEO:Region}
and Prabhakaran, Tse, and
Ramchandran~\cite{Prabhakaran:ISIT04}
is a consequence of the outer bound provided in this paper. This
does not imply, however, that the task of 
proving the converse result is made 
any easier by
our bound. In fact, comparing Appendix~\ref{Gaussian:proof} to the
original works shows that proving Proposition~\ref{Gaussian:Prop}
is as formidable a task as proving the converse result unaided. But this
is still an improvement over the Berger-Tung outer bound, the closure
of which
we show in Appendix~\ref{BT:loose:Gaussian} contains points outside
the rate-distortion region.

We end this section by mentioning that Oohama's~\cite{Oohama:CEO:Region}
converse
is actually more general than the result described here,
in that Oohama permits one of the encoders to make noise-free
observations (i.e., $\sigma^2_1 = 0$). Comparing
Oohama's proof to Appendix~\ref{Gaussian:proof} shows that
the outer bound supplied in this paper also recovers this more general result.

\section{Recovery of Discrete Converse Results}
\label{recover}

Having seen that the new outer bound recovers the converse 
of
Oohama~\cite{Oohama:CEO:Region} and Prabhakaran, Tse, and 
Ramchandran~\cite{Prabhakaran:ISIT04} for the Gaussian CEO
problem, we show in this final section
that it also recovers the converse results for the
discrete problems of 
Slepian and Wolf~\cite{Slepian:SW}, 
Wyner~\cite{Wyner:SCSI},
Ahlswede and K\"{o}rner~\cite{Ahlswede:SCSI}, 
Wyner and Ziv~\cite{Wyner:WZ},
Gel`fand and Pinsker~\cite{Gelfand:CEO}, 
Berger and 
Yeung~\cite{Berger:BY}, and 
Gastpar~\cite{Gastpar:WZ}.
The outer bound also recovers the converse
result for the problem studied by
K\"{o}rner and Marton~\cite{Korner:ModTwo}, 
although the proof of this fact is not as
interesting. We shall therefore
focus on the others.
To recover these converse results, we shall use the following
conclusive result for a special case of the problem.

Suppose that there exists a function $g : \mathcal{Y}_0 \mapsto
\mathcal{Y}_0$ such that $Y_1,\ldots,Y_L$ are conditionally 
independent given $(g(Y_0),Y_{L+1})$. Also let 
$\mathcal{Z}_1 = \mathcal{Y}_0$ 
and \begin{equation*}
d_1(Y_0,\mathbf{Y},Y_{L+1},Z_1) = 
\begin{cases}
0 & \text{if $Z_1 = Y_0$} \\
1 & \text{otherwise.}
\end{cases}
\end{equation*}
We make no other assumptions about the problem.
We would like to characterize the set $\overline{\mathcal{RD}_\star} 
\cap \{D_1 = 0\}$.
In words, conditioned on the side information and
some function of the hidden variable,
the observations are independent, and the hidden variable must
be reproduced losslessly. Note that $\overline{\mathcal{RD}_\star}
\cap \{D_1 = 0\}$ will be empty unless $H(Y_0|\mathbf{Y},Y_{L+1}) = 0$.
Gel`fand and Pinsker~\cite{Gelfand:CEO} refer to this
condition as ``completeness of observations.''
\begin{prop}
\label{reduction}
For this problem, 
\begin{align}
\label{recover:main}
\overline{\mathcal{RD}_o} \cap \{D_1 = 0\} & = 
\overline{\mathcal{RD}_i^{BT}} \cap \{D_1 = 0\} = 
\overline{\mathcal{RD}_\star} \cap \{D_1 = 0\} \\
\label{recover:region}
  & = \bigcup_{\gamma \in \Gamma_i^{BT}} \mathcal{RD}_i^{BT}(\gamma)
   \cap \{D_1 = 0\}.
\end{align}
\end{prop}

\emph{Proof.}
To show~(\ref{recover:main}),
it suffices to show that $\overline{\mathcal{RD}_o} \cap \{D_1 = 0\}$ is
contained in $\overline{\mathcal{RD}_i^{BT}} \cap \{D_1 = 0\}$.
Suppose $(\mathbf{R},\epsilon,D_2,\ldots,D_K)$ is a point in
$\mathcal{RD}_o$ and $\epsilon \le 1/2$. 
By choosing $X = g(Y_0)$ in Definition~\ref{Me:defn},
we see that there exists 
$(\mathbf{U},\mathbf{Z},W,\tilde{T})$ in $\Gamma_o$
such that
\begin{align}
\nonumber
\Pr(Z_1 \ne Y_0) & \le \epsilon  \\
\nonumber
E[d_k(Y_0,\mathbf{Y},Y_{L+1},Z_k)] & \le D_k \ \text{for all $k \ge 2$},
\end{align}
and for all $A$,
\begin{equation*}
\sum_{\ell \in A} R_\ell \ge 
     I(g(Y_0);\mathbf{U}_A|\mathbf{U}_{A^c},Y_{L+1},\tilde{T}) +
\sum_{\ell \in A} I(Y_\ell;U_\ell|g(Y_0),Y_{L+1},W,\tilde{T}).
\end{equation*}
Now
\begin{align*}
I(g(Y_0);\mathbf{U}_A|\mathbf{U}_{A^c},Y_{L+1},\tilde{T}) & = 
    H(g(Y_0)|\mathbf{U}_{A^c},Y_{L+1},\tilde{T}) \\
 &  \phantom{H(g(Y_0)|}  
   \mbox{} - H(g(Y_0)|\mathbf{U},Y_{L+1},\tilde{T}) \\
 & \ge H(g(Y_0)|\mathbf{U}_{A^c},Y_{L+1},W,\tilde{T}) - H(g(Y_0)|g(Z_1)),
\end{align*}
where we have used the fact that
\begin{equation*}
g(Y_0) \markov Y_0 \markov (\mathbf{U},Y_{L+1},\tilde{T}) \markov Z_1
  \markov g(Z_1).
\end{equation*}
By Fano's inequality~\cite[Lemma~1.3.8]{CK:IT},
\begin{equation*}
H(g(Y_0)|g(Z_1)) \le h(\epsilon) + \epsilon \log(|\mathcal{Y}_0|).
\end{equation*}
Thus
\begin{align*}
I(g(Y_0);\mathbf{U}_A|\mathbf{U}_{A^c},Y_{L+1},\tilde{T}) & \ge
H(g(Y_0)|\mathbf{U}_{A^c},Y_{L+1},W,\tilde{T}) - 
     h(\epsilon) - \epsilon \log(|\mathcal{Y}_0|) \\
& \ge I(g(Y_0);\mathbf{U}_A|\mathbf{U}_{A^c},Y_{L+1},W,\tilde{T})  \\
   & \phantom{\ge I(g(Y_0);}  - h(\epsilon) - \epsilon \log(|\mathcal{Y}_0|). \\
\end{align*}
It follows that
\begin{align*}
\sum_{\ell \in A} [R_\ell + h(\epsilon) + 
\epsilon\log(|\mathcal{Y}_0|) ] & \ge
I(g(Y_0);\mathbf{U}_A|\mathbf{U}_{A^c},Y_{L+1},W,\tilde{T})  \\
 & \phantom{I(Y_0;\mathbf{U}_A)} + 
       \sum_{\ell \in A} I(Y_\ell;U_\ell|g(Y_0),Y_{L+1},W,\tilde{T}) \\
 & = I(g(Y_0);\mathbf{U}_A|\mathbf{U}_{A^c},Y_{L+1},W,\tilde{T})  \\
 & \phantom{I(Y_0;\mathbf{U}_A|} + 
   I(\mathbf{Y}_A;\mathbf{U}_A|\mathbf{U}_{A^c},g(Y_0),Y_{L+1},W,\tilde{T}) \\
 & = I(g(Y_0),\mathbf{Y}_A;\mathbf{U}_A|\mathbf{U}_{A^c},Y_{L+1},W,\tilde{T}) \\
 & \ge I(\mathbf{Y}_A;\mathbf{U}_A|\mathbf{U}_{A^c},Y_{L+1},W,\tilde{T}) .
\end{align*}
If we now define $T = (W,\tilde{T})$, it is evident that 
$(\mathbf{U},\mathbf{Z},T)$ is in $\Gamma_i^{BT}$ and the point
\begin{equation*}
(R_1 + h(\epsilon) + \epsilon\log(|\mathcal{Y}_0|),
 \ldots, R_L + h(\epsilon) + \epsilon\log(|\mathcal{Y}_0|),\epsilon,
D_2,\ldots,D_K)
\end{equation*}
is in $\mathcal{RD}_i^{BT}$. This implies that 
\begin{equation*}
\overline{\mathcal{RD}_o} \cap \{D_1 = 0\}
  \subset \overline{\mathcal{RD}_i^{BT}} \cap \{D_1 = 0\},
\end{equation*}
which proves~(\ref{recover:main}). To prove~(\ref{recover:region}),
it suffices to show that $\mathcal{RD}_i^{BT}$ is closed. This
is shown in Appendix~\ref{closed}. \QED

The differences between this result and that of Gel`fand and
Pinsker~\cite{Gelfand:CEO} are numerous but minor.
The most visible differences are that Gel`fand and Pinsker's
model does not allow for side information at the decoder or
distortion constraints beyond the one on $Y_0$. 
Indeed, the region given here reduces to theirs
when these extensions are ignored. 
Thus this result
seems to be a generalization of theirs, albeit a trivial one
since their proof can be modified to handle these extensions. 
A closer comparison,
however, reveals that they define the rate region more
stringently than we do here.
Thus, our result does not recover theirs, strictly speaking,
although it does recover 
the converse component of their result since our definitions
are weaker.

The reason for including side information and
additional distortion constraints in the model is that they
enable us to also recover the converse results for
the other problems mentioned earlier.
For instance, Gastpar~\cite{Gastpar:WZ} considers the problem
of reproducing the observations individually, subject to
separate distortion constraints, under the assumption
that the decoder is provided with side information
that makes the observations conditionally independent.
His converse result can be recovered by setting
$Y_0 = Y_{L+1}$. It is easily
verified that, under this condition, our region
coincides with his. The classical Wyner-Ziv 
problem~\cite{Wyner:WZ} can be viewed as Gastpar's problem 
with a single encoder ($L=1$). So that converse result is 
recovered too.

Berger and Yeung~\cite{Berger:BY} solve the two-encoder
problem in which the observations are to be reproduced 
individually, with at least one of the two being reproduced
losslessly. In our notation, this corresponds to setting
$L = 2$ and $Y_1 = Y_0$. Note that our
conditional independence assumption necessarily holds
in this case.

To see that under these assumptions, our region reduces
to theirs, suppose $(R_1,R_2,D_2) \in \mathcal{RD}_i^{BT}(\gamma)
\cap \{D_1 = 0\}$ for some $\gamma \in \Gamma_i^{BT}$. Then
\begin{equation}
\label{BY:1}
R_1 \ge I(Y_1;U_1|U_2,T) = H(Y_1|U_2,T).
\end{equation}
Also,
\begin{align}
\nonumber
R_2 \ge I(Y_2;U_2|U_1,T) & \ge I(Y_2;U_2|U_1,Y_1,T) \\
\label{BY:2}
 & = I(Y_2;U_2|Y_1,T),
\end{align}
where we have used the fact that 
\begin{equation*}
U_2 \markov (Y_2,U_1,T) \markov Y_1
\end{equation*}
(see Cover and Thomas~\cite[p.~33]{Cover:IT}). Finally,
\begin{align}
\nonumber
R_1 + R_2 & \ge I(Y_1,Y_2;U_1,U_2|T) \\
\nonumber
   & \ge I(Y_1;U_1,U_2|T) + I(Y_2;U_1,U_2|Y_1,T) \\
\label{BY:3}
   & \ge H(Y_1) + I(Y_2;U_2|Y_1,T).
\end{align}
It is now evident that the two regions are 
identical (c.f.~\cite[p.~230]{Berger:BY}). Thus the
converse result of Berger and Yeung is a consequence of the
outer bound provided here.

The classical problem of source coding with side 
information~\cite{Wyner:SCSI,Ahlswede:SCSI} can be viewed as a
special case of the Berger-Yeung problem in which $D_2$ exceeds
the maximum value of $d_2$, the distortion measure for $Y_2$.
Berger and Yeung demonstrate how, under this assumption, the
region described above reduces
to the one given by Wyner~\cite{Wyner:SCSI} and 
Ahlswede and K\"{o}rner~\cite{Ahlswede:SCSI}.
\emph{Ipso facto}, the converse result for this problem is also recovered.

This paper ends the way it began, with the result
of Slepian and Wolf~\cite{Slepian:SW}.
Here the aim is to losslessly reproduce all of the observations.
For two encoders ($L = 2$), this can be
viewed as a special case of the problem of Berger and Yeung.
These authors show how the region described in 
Eqs.~(\ref{BY:1})--(\ref{BY:3}) reduces to the one 
given at the beginning of the paper. The result for
more than two encoders can be viewed as a special case of 
Proposition~\ref{reduction} in which $Y_0 = \mathbf{Y}$.
In this case, if $\mathbf{R} \in 
\overline{\mathcal{RD}_o} \cap \{D_1 = 0\}$, then
for any $A$,
\begin{align*}
\sum_{\ell \in A} R_\ell & \ge 
  I(\mathbf{Y}_A;\mathbf{U}_A|\mathbf{U}_{A^c},T) \\
  & = H(\mathbf{Y}_A|\mathbf{U}_{A^c},T) \\
  & \ge H(\mathbf{Y}_A|\mathbf{Y}_{A^c},T),
\end{align*}
since $\mathbf{Y}_A \markov (\mathbf{Y}_{A^c},T) \markov (\mathbf{U}_{A^c},T)$.
Now $\mathbf{Y}$ is independent of $T$, so
\begin{equation*}
\sum_{\ell \in A} R_\ell \ge H(\mathbf{Y}_A|\mathbf{Y}_{A^c}),
\end{equation*}
which is the well-known rate region for this problem.
Thus the converse of Slepian and Wolf is also recovered.
For this result, as with the others, our outer bound
dispenses with the need to prove a custom converse coding
theorem. In fact, Proposition~\ref{reduction} can
be viewed as unifying all of the results
in this discussion,
assuming one is willing to ignore the discrepancies in
the definition of the rate-distortion region
mentioned earlier.

\appendix

\section{Sum-Rate Achievability for the Binary Erasure CEO Problem}
\label{BECEO:achieve}

Showing that a particular rate-distortion vector
is achievable using the Berger-Tung inner bound is mostly
a matter of finding the proper ``test channels''
$Y_\ell \rightarrow U_\ell$ for the encoders. To prove
(\ref{BECEO:sumrate}), we use binary erasure test
channels that are identically distributed across
the encoders. In this appendix and the next two,
the notation is drawn from Section~\ref{BECEO:subsection}.

\begin{lemma}
For any $p^L \le D \le 1$,
\begin{multline*}
\lim_{\lambda \rightarrow \infty} \mathcal{R}_i^{BT}(D,\lambda)
 \le (1-D)\log 2
\mbox{} + L\left[h\left(D^{1/L}\right) -
(1-p)h\left(\frac{D^{1/L} - p}{1-p}\right)\right].
\end{multline*}
\end{lemma}
\emph{Proof.} Fix $D$ and let 
$\tilde{N}_1,\ldots,\tilde{N}_L$ be i.i.d., independent
of $Y_0,\ldots,Y_L$, with
\begin{align*}
\Pr\left(\tilde{N}_1 = 0\right) & = \frac{D^{1/L} - p}{1-p} \\
\Pr\left(\tilde{N}_1 = 1\right) & = 1 - \frac{D^{1/L} - p}{1-p} .
\end{align*}
For $\ell$ in $\{1,\ldots,L\}$, let $U_\ell = Y_\ell \cdot \tilde{N}_i$. Then
let
\begin{equation*}
Z_1 = \sgn\left(\sum_{\ell = 1}^L U_\ell \right) := 
\begin{cases}
-1 & \text{if} \ \sum_{\ell = 1}^L U_\ell < 0 \\
0 & \text{if} \ \sum_{\ell = 1}^L U_\ell = 0 \\
1 & \text{otherwise}.
\end{cases}
\end{equation*}
Then for all $\lambda$,
\begin{equation*}
E\left[d^\lambda_1(Y_0,\mathbf{Y},Z_1)\right] = 
  \Pr(U_\ell = 0 \ \text{for all} \ \ell)
  = \left[\Pr(U_1 = 0)\right]^L = D.
\end{equation*}
Thus $(\mathbf{R},D)$ is contained
in $\mathcal{RD}_i^{BT}$ for all $\lambda$ if for all $A \subset
\{1,\ldots,L\}$,
\begin{equation}
\label{contrapoly}
\sum_{\ell \in A} R_\ell \ge I(\mathbf{Y}_A;\mathbf{U}_A|\mathbf{U}_{A^c}).
\end{equation}
The rate vectors satisfying this collection of inequalities are
known to form a contrapolymatroid~\cite{Viswanath:DIMACS:Source,
Chen:CEO:Bounds}. As such, 
there exist rate vectors $\mathbf{R}$ satisfying~(\ref{contrapoly})
such that
\begin{equation*}
\sum_{\ell = 1}^L R_\ell = I(\mathbf{Y};\mathbf{U}).
\end{equation*}
In particular, this holds for any vertex 
of~(\ref{contrapoly})~\cite{Viswanath:DIMACS:Source,Chen:CEO:Bounds}.
Now
\begin{align*}
I(\mathbf{Y};\mathbf{U}) & = I(Y_0,\mathbf{Y};\mathbf{U}) \\
   & = I(Y_0;\mathbf{U}) + I(\mathbf{Y};\mathbf{U}|Y_0) \\
   & = I(Y_0;\mathbf{U}) + L \; I(Y_1;U_1|Y_0).
\end{align*}
But $I(Y_0;\mathbf{U}) = (1-D)\log 2$ and
\begin{align*}
I(Y_1;U_1|Y_0) & = H(U_1|Y_0) - H(U_1|Y_1) \\
    & = h(D^{1/L}) - (1-p)h\left(\frac{D^{1/L} - p}{1-p}\right).
\end{align*}
Then for any $\lambda$, there exist vectors $(\mathbf{R},D)$ in
$\mathcal{RD}_i^{BT}(\lambda)$ such that
\begin{equation*}
\sum_{\ell = 1}^L R_\ell = (1-D)\log 2 + L\left[
  h(D^{1/L}) - (1-p)h\left(\frac{D^{1/L} - p}{1-p}\right) \right].
\end{equation*}
The conclusion follows. \QED

\section{Sum-Rate Converse for the Binary Erasure CEO Problem}
\label{BECEO:converse}

We evaluate the outer bound's sum-rate constraint for the
binary erasure CEO problem via a sequence of lemmas. Throughout
this appendix, $g(\cdot)$ will denote the function on $[p,\infty)$
defined by
\begin{equation*}
g(x) = \begin{cases}
h(x) - (1-p) h(\frac{x - p}{1 - p}) & p \le x \le 1 \\
0 & x > 1.
\end{cases}
\end{equation*}
We begin by proving several facts about $g(\cdot)$. For this, the following
calculations are useful.
\begin{lemma}
\label{calc}
For all $x$ in $(\log p, 0]$,
\begin{equation}
\label{firstcalc}
e^x \log(e^x - p) - x e^x \le - p
\end{equation}
and
\begin{equation}
\label{secondcalc}
e^x \log(e^x - p) - e^x (x+1) + \frac{e^{2x}}{e^x - p} \ge 0.
\end{equation}
\end{lemma}
\emph{Proof.} 
It is well known that
\begin{equation*}
\log\left(\frac{1}{1 - z}\right) \ge z \ \text{for all} \ z < 1.
\end{equation*}
Replacing $z$ with $p e^{-x}$ and rearranging yields~(\ref{firstcalc}).
To see~(\ref{secondcalc}),
note that~(\ref{firstcalc})
implies that the first derivative of 
\begin{equation}
\label{calcfun}
(e^x - p)\log(e^x - p) - (e^x - p)(x+1) + e^x
\end{equation}
is nonpositive on $(\log p,0]$. Since the function 
in~(\ref{calcfun}) is nonnegative at $x = 0$, it follows 
that 
\begin{equation}
(e^x - p)\log(e^x - p) - (e^x - p)(x+1) + e^x \ge 0
\end{equation}
for all $x$ in $(\log p,0]$.
One can now obtain~(\ref{secondcalc}) by multiplying
both sides by $e^x$ and dividing both sides by $(e^x - p)$.
\QED
\begin{lemma}
\label{convex}
The function $g(e^x)$ is nonincreasing and convex as a function
of $x$ on $[\log p, \infty)$.
\end{lemma}
\emph{Proof.} The first derivative of $g(e^x)$ on $(\log p,0)$ is
\begin{equation*}
e^x \log (e^x - p) - x e^x.
\end{equation*}
This observation, the first conclusion of 
Lemma~\ref{calc}, and the continuity of $g(\cdot)$ together
imply that $g(e^x)$ is nonincreasing
on $[\log p, 0]$. Since $g(e^x)$ is constant on 
$[0,\infty)$, it follows that $g(e^x)$ is nonincreasing
on $[\log p, \infty)$. The second derivative of $g(e^x)$
on $(\log p,0)$ is 
\begin{equation*}
e^x \log(e^x - p) - e^x (x+1) + \frac{e^{2x}}{e^x - p}.
\end{equation*}
This observation, the second conclusion of Lemma~\ref{calc},
and the continuity of $g(\cdot)$ together imply
that $g(e^x)$ is convex on $[\log p, 0]$. Since
$g(e^x)$ is nonincreasing on $[\log p,\infty)$ and
constant on $[0,\infty)$, it follows that
$g(e^x)$ is convex on $[\log p, \infty)$. \\
\mbox{} \hfill \QED

\begin{cor}
\label{convexcor}
The function
$g(y^{1/L})$ is nonincreasing and convex in $y$ on
$[p^L,\infty)$.
\end{cor}
\emph{Proof.} $g(y^{1/L}) = g(e^x)$ with $x = (1/L)\log y$,
and $g(e^x)$ is convex and nonincreasing while $(1/L)\log(\cdot)$ is
concave and nondecreasing. \QED

The next lemma is central to our evaluation of the
outer bound's sum rate. Note that condition (\emph{i})
in the hypothesis implies that 
$\Pr(Y_0 \cdot Z_1 < 0) = 0$. That is, the reproduction
$Z_1$ is never in error (although it may be an erasure).
\begin{lemma}
\label{central}
Suppose $p^L \le D$ and $(\mathbf{U},Z_1)$ is
such that
\begin{enumerate}
\item[(\emph{i})] $E[d_1^\lambda(Y_0,Z_1)] \le D \ \text{for all} \ \lambda$,
\item[(\emph{ii})] $U_\ell \markov Y_\ell \markov
     (Y_0,\mathbf{Y}_{\ell^c},\mathbf{U}_{\ell^c})$
  for all $\ell$, and
\item[(\emph{iii})] $(Y_0,\mathbf{Y}) \markov \mathbf{U} \markov Z_1$.
\end{enumerate}
Then
\begin{equation*}
\frac{1}{L} \sum_{\ell = 1}^L I(Y_\ell;U_\ell|Y_0) \ge g(D^{1/L}).
\end{equation*}
\end{lemma}

\emph{Proof.} 
For each encoder $\ell$, let
\begin{align*}
A_{\ell,+} & = \{u \in \mathcal{U}_\ell : \Pr(U_\ell = u|Y_0 = 1) > 0\} \\
A_{\ell,-} & = \{u \in \mathcal{U}_\ell : \Pr(U_\ell = u|Y_0 = -1) > 0\}.
\end{align*}
Then define
\begin{equation*}
\tilde{U}_\ell = \begin{cases}
1  & \text{if $U_\ell \in A_{\ell,+} \backslash A_{\ell,-}$} \\
-1  & \text{if $U_\ell \in A_{\ell,-} \backslash A_{\ell,+}$} \\
0  &  \text{otherwise}.
\end{cases}
\end{equation*}
Finally, let
\begin{align*}
\delta_{\ell,+} & = \Pr(\tilde{U}_\ell = 0|Y_\ell = 1) \\
\delta_{\ell,-} & = \Pr(\tilde{U}_\ell = 0|Y_\ell = -1).
\end{align*}
Then
\begin{align*}
\frac{1}{L} \sum_{\ell = 1}^L I(Y_\ell;U_\ell|Y_0) & \ge \frac{1}{L} 
   \sum_{\ell = 1}^L I(Y_\ell;\tilde{U}_\ell|Y_0) \\
  & = \frac{1}{L} \sum_{\ell = 1}^L \left[H(\tilde{U}_\ell|Y_0) - 
        H(\tilde{U}_\ell|Y_\ell)\right] \\
  & = \frac{1}{L} \sum_{\ell = 1}^L \Bigg[ \frac{1}{2}h\left(p
    + (1-p)\delta_{\ell,+} \right) + \frac{1}{2}h\left(p 
    + (1-p)\delta_{\ell,-} \right) \\
  & \phantom{= \frac{1}{L} \sum_{\ell = 1}^L \Bigg[} - \frac{1}{2}(1-p)
    h(\delta_{\ell,+}) - \frac{1}{2}(1-p) h(\delta_{\ell,-})\Bigg].
\end{align*}
Since $Y_0 \cdot Z_1 \ge 0$ a.s., on the event $Z_1 = 1$ 
we must have $Y_0 = 1$ and hence $U_\ell \in A_{\ell,+}$ for all $\ell$.
In addition, the condition $Y_0 \markov \mathbf{U} \markov Z_1$ dictates that
when $Z_1 = 1$ we must have 
$U_\ell \in A_{\ell,+} \backslash A_{\ell,-}$
for some $\ell$,
for otherwise we would have $\Pr(Y_0 \cdot Z_1 = -1) > 0$. All of
this implies that $\sgn(\sum_{\ell = 1}^L \tilde{U}_\ell) = 1$ on
the event that $Z_1 = 1$.
Similarly, $\sgn(\sum_{\ell = 1}^L \tilde{U}_\ell) = -1$ on
the event $Z_1 = -1$. Thus
$\sgn(\sum_{\ell = 1}^L \tilde{U}_\ell) = 0$ implies that
$Z_1 = 0$, so
\begin{equation*}
   \Pr\left(\sgn\left(\sum_{\ell = 1}^L \tilde{U}_\ell\right) = 0\right)
    \le \Pr(Z_1 = 0) = D.
\end{equation*}
This implies that
\begin{equation*}
\frac{1}{2} \prod_{\ell = 1}^L (p + (1-p) \delta_{\ell,+})
 + \frac{1}{2} \prod_{\ell = 1}^L (p + (1-p) \delta_{\ell,-}) \le D.
\end{equation*}
Thus
\begin{align*}
\frac{1}{L} \sum_{\ell = 1}^L I(Y_\ell;U_\ell|Y_0) & 
  \ge \inf\Bigg\{ \frac{1}{L} \sum_{\ell = 1}^L 
  \frac{1}{2} \Big[h(p + (1-p)\delta_{\ell,+}) -
  (1-p) h(\delta_{\ell,+}) \\
  & \phantom{\inf\Big\{ \sum_{\ell = 1}^L \frac{1}{2} \Big[}
   + h(p + (1-p)\delta_{\ell,-}) -
  (1-p) h(\delta_{\ell,-}) \Big] : \\
 & \phantom{\inf\Big\{ \sum_{\ell = 1}^L \frac{1}{2} \Big[}
  \delta_{\ell,+}, \delta_{\ell,-} \in [0,1] \ \text{for all} \ 
    \ell \ \text{and} \\
 & \phantom{\inf\Big\{ }
 \frac{1}{2} \prod_{\ell = 1}^L (p + (1-p) \delta_{\ell,+})
 + \frac{1}{2} \prod_{\ell = 1}^L (p + (1-p) \delta_{\ell,-}) \le D
  \Bigg\}.
\end{align*}
This optimization problem is not convex, but if we 
change variables to
\begin{align*}
\Delta_{\ell,+} & = \log(p + (1-p) \delta_{\ell,+}) \\
\Delta_{\ell,-} & = \log(p + (1-p) \delta_{\ell,-}),
\end{align*}
then it can be rewritten as
\begin{align*}
 & \inf\Bigg\{ \frac{1}{L} \sum_{\ell = 1}^L 
  \frac{1}{2} \Bigg[h\left(e^{\Delta_{\ell,+}}\right) -
  (1-p) h\left(\frac{e^{\Delta_{\ell,+}} - p}{1 - p}\right) \\
  & \phantom{\ge \inf\Big\{ } + h\left(e^{\Delta_{\ell,-}}\right) -
  (1-p) h\left(\frac{e^{\Delta_{\ell,-}} - p}{1 - p}\right) 
  \Bigg] : \\
  & \phantom{\ge \inf\Big\{ } \Delta_{\ell,+}, \Delta_{\ell,-} \in
  [\log p,0] \ \text{for all} \ \ell \ \text{and} \\
& \phantom{\ge \inf\Big\{ } \frac{1}{2} \exp\left(\sum_{\ell = 1}^L 
    \Delta_{\ell,+}\right)
 + \frac{1}{2} \exp\left(\sum_{\ell = 1}^L \Delta_{\ell,-}\right) \le D 
  \Bigg\} \\
& = \inf\Bigg\{ \frac{1}{L} \sum_{\ell = 1}^L
  \frac{1}{2} \left[g\left(e^{\Delta_{\ell,+}}\right) + 
     g\left(e^{\Delta_{\ell,-}}\right) \right] :  \\
  & \phantom{\ge \inf\Big\{ }  \Delta_{\ell,+}, \Delta_{\ell,-} \in
  [\log p,0] \ \text{for all} \ \ell \ \text{and} \\
& \phantom{\ge \inf\Big\{ } \frac{1}{2} \exp\left(\sum_{\ell = 1}^L 
   \Delta_{\ell,+}\right)
 + \frac{1}{2} \exp\left(\sum_{\ell = 1}^L \Delta_{\ell,-}\right) \le D
  \Bigg\},
\end{align*}
which is convex by Lemma~\ref{convex}. Thus
we may assume without loss of optimality that
\begin{equation*}
\Delta_{1,+} = \Delta_{2,+} = \cdots = \Delta_{L,+} =: \Delta_{+}
\end{equation*}
and
\begin{equation*}
\Delta_{1,-} = \Delta_{2,-} = \cdots = \Delta_{L,-} =: \Delta_{-}.
\end{equation*}
This gives
\begin{align*}
\frac{1}{L} \sum_{\ell = 1}^L I(Y_\ell;U_\ell|Y_0) & \ge \inf\Bigg\{\frac{1}{2}
    g\left(e^{\Delta_+}\right) + \frac{1}{2} g\left(e^{\Delta_-}\right) :
   \Delta_+, \Delta_- \in [\log p, 0] : \\
 & \phantom{\ge \inf\Bigg\{}
  \frac{1}{2}e^{L \Delta_+} + \frac{1}{2}e^{L \Delta_-} 
             \le D\Bigg\} \\
  & \ge \inf\Big\{g\left(e^\Delta\right) : \Delta \in [\log p, 0] :
   e^{L\Delta} \le D\Big\} \\
  & \ge g(D^{1/L}),
\end{align*}
by Lemma~\ref{convex}. \QED

The quantity $I(Y_\ell;U_\ell|Y_0)$ can
be interpreted as the amount of information that
the $\ell$th encoder sends about its observation 
\emph{noise}\footnote{This terminology is due
to Prabhakaran, Tse, and Ramchandran.}. Lemma~\ref{central} then
says that if a fraction $D$ of the output symbols is
allowed to be erased and no errors are allowed,
then the amount of information that the
average encoder must send about its observation noise is
at least $g(D^{1/L})$. We would like to extend this last
assertion to allow ``few'' decoding errors instead of none.
To this end, we will employ the following cardinality bound
on the alphabet sizes of the auxiliary random 
variables $U_1,\ldots,U_L$.
\begin{lemma}
\label{cardinality}
Let $(\mathbf{U},Z_1)$ be such that
\begin{enumerate}
\item[(\emph{i})] $U_\ell \markov Y_\ell \markov
   (Y_0,\mathbf{Y}_{\ell^c},\mathbf{U}_{\ell^c})$ 
  for all $\ell$, and
\item[(\emph{ii})] $(Y_0,\mathbf{Y}) \markov \mathbf{U} \markov Z_1$.
\end{enumerate}
Then for any $\lambda$, there exist
alternate random variables $\tilde{\mathbf{U}}$ and
$\tilde{Z_1}$ also satisfying
(\emph{i}) and (\emph{ii}) such that
\begin{align*}
E[d_1^\lambda(Y_0,\tilde{Z_1})] & \le E[d_1^\lambda(Y_0,Z_1)], \\
I(Y_\ell;\tilde{U}_\ell|Y_0) & = I(Y_\ell;U_\ell|Y_0) \ \text{for all $\ell$},
\end{align*}
and
\begin{equation*}
|\mathcal{U}_\ell| \le |\mathcal{Y}_\ell| + 1 \ \text{for all $\ell$}.
\end{equation*}
\end{lemma}
See Wyner and Ziv~\cite[Theorem~A2]{Wyner:WZ} or Csisz\'{a}r and 
K\"{o}rner~\cite[Theorem~3.4.6]{CK:IT}
for proofs of similar results.
The next lemma is the desired extension of Lemma~\ref{central}.
\begin{lemma}
\label{continuity}
Suppose $p^L \le D$ and $(\mathbf{U},Z_1)$ is such that
\begin{enumerate}
\item[(\emph{i})] $E[d_1^\lambda(Y_0,Z_1)] \le D$,
\item[(\emph{ii})] $U_\ell \markov Y_\ell \markov (Y_0,\mathbf{Y}_{\ell^c},
   \mathbf{U}_{\ell^c})$ for all $\ell$, and
\item[(\emph{iii})] $(Y_0,\mathbf{Y}) \markov \mathbf{U} \markov Z_1$.
\end{enumerate}
If 
\begin{equation*}
\frac{32L}{p(1-p)} \left(\frac{2D}{\lambda}\right)^{1/L} \le \delta
  \le \frac{1}{2},
\end{equation*}
then
\begin{equation*}
\frac{1}{L} \sum_{\ell = 1}^L I(Y_\ell;U_\ell|Y_0) \ge 
  g\left((D+ \delta)^{1/L}\right)
   + 2\delta \log\frac{\delta}{5}.
\end{equation*}
\end{lemma}
\emph{Proof.} 
By Lemma~\ref{cardinality}, we may assume that 
$\mathcal{U}_\ell = \{1,\ldots,4\}$ for each $\ell$. We may
also assume that $Z_1$ is a deterministic function of
$\mathbf{U}$: $Z_1 = \phi(\mathbf{U})$. Define
\begin{align*}
A_{\ell,+} & = \Bigg\{u_\ell \in \mathcal{U}_\ell : 
    \exists \ \mathbf{u}_{\ell^c} :
     \phi(\mathbf{u}) = 1 \ \text{and} \\
   & \phantom{\Bigg\{u_\ell \in \mathcal{U}_\ell : }
   \Pr(U_\ell = u_\ell|Y_0 = -1) 
    = \min_{j \in \{1,\ldots,L\}} \Pr(U_j = u_j|Y_0 = -1) \Bigg\} \\
A_{\ell,-} & = \Bigg\{u_\ell \in \mathcal{U}_\ell : \exists \ 
      \mathbf{u}_{\ell^c} :
     \phi(\mathbf{u}) = -1 \ \text{and} \\
   & \phantom{\Bigg\{u_\ell \in \mathcal{U}_\ell : }
    \Pr(U_\ell = u_\ell|Y_0 = 1)
    = \min_{j \in \{1,\ldots,L\}} \Pr(U_j = u_j|Y_0 = 1) \Bigg\}.
\end{align*}
We now define random variables $(\tilde{\mathbf{U}},\tilde{\mathbf{Z}})$
to replace $(\mathbf{U},\mathbf{Z})$.
The replacements will be close to the originals in distribution
but will have the property that $\Pr(Y_1\cdot\tilde{Z}_1 < 0) = 0$.
That is, $\tilde{Z}_1$ will never be in error.
Set $\mathcal{\tilde{U}}_\ell = \{1,\ldots,5\}$ for each $\ell$, and
let
\begin{align*}
\Pr(\tilde{U}_\ell = i|Y_\ell = 1) & = 
\begin{cases}
 0 & \text{if $i \in A_{\ell,-}$} \\
\Pr(U_\ell \in A_{\ell,-}|Y_\ell = 1) & \text{if $i = 5$} \\
\Pr(U_\ell = i|Y_\ell = 1) & \text{otherwise} \\
\end{cases} \\
\Pr(\tilde{U}_\ell = i|Y_\ell = -1) & = 
\begin{cases}
0  & \text{if $i \in A_{\ell,+}$} \\
\Pr(U_\ell \in A_{\ell,+}|Y_\ell = -1) & \text{if $i = 5$} \\
\Pr(U_\ell = i|Y_\ell = -1) & \text{otherwise} \\
\end{cases} \\
\Pr(\tilde{U}_\ell = i|Y_\ell = 0) & = 
\begin{cases}
0  & \text{if $i \in A_{\ell,+} \cup A_{\ell,-}$} \\
\Pr(U_\ell \in A_{\ell,+} \cup A_{\ell,-}|Y_\ell = 0) & \text{if $i = 5$} \\
\Pr(U_\ell = i|Y_\ell = 0) & \text{otherwise}.
\end{cases}
\end{align*}
Then define
\begin{equation*}
\tilde{Z}_1 = \tilde{\phi}(\tilde{\mathbf{U}}) := \begin{cases}
\phi(\tilde{u}) & \text{if $\tilde{u}_\ell \le 4$ for all $\ell$} \\
 0 & \text{otherwise}.
\end{cases}
\end{equation*}
There is a natural way of coupling $\mathbf{\tilde{U}}$ to
$\mathbf{U}$ such that if $\tilde{U}_\ell$ is in
$\{1,\ldots,4\}$ then $\tilde{U}_\ell = U_\ell$.
With this coupling in mind, it is evident that
\begin{align*}
E[d^\lambda_1(Y_0,\tilde{Z}_1)] & = 
   E[d_1^\lambda(Y_0,\tilde{Z}_1)1(\max(\tilde{U}_1,\ldots,
    \tilde{U}_L) \le 4)] \\
    & \phantom{E[} + 
   E[d_1^\lambda(Y_0,\tilde{Z}_1)1(\max(\tilde{U}_1,\ldots,\tilde{U}_L) = 5)] \\
   & \le E[d_1^\lambda(Y_0,Z_1)1(\max(\tilde{U}_1,\ldots,
    \tilde{U}_L) \le 4)] \\
   & \phantom{E[} + \Pr(\max(\tilde{U}_1,\ldots,\tilde{U}_L) = 5) \\
   & \le D + \Pr(\max(\tilde{U}_1,\ldots,\tilde{U}_L) = 5).
\end{align*}
Now for any $\ell$ in $\{1,\ldots,L\}$,
\begin{align*}
\Pr(\tilde{U}_\ell = 5) & = 
    \frac{1-p}{2} \Pr(U_\ell \in A_{\ell,-}|Y_\ell = 1) +
   \frac{1-p}{2} \Pr(U_\ell \in A_{\ell,+}|Y_\ell = -1) \\
  &  \phantom{=} + p \Pr(U_\ell \in A_{\ell,+} \cup A_{\ell,-}|Y_\ell = 0).
\end{align*}
By the union bound, this is upper bounded by
\begin{multline*}
  \left[\frac{1-p}{2} \Pr(U_\ell \in A_{\ell,-}|Y_\ell = 1) +
        p \Pr(U_\ell \in A_{\ell,-}|Y_\ell = 0)\right]  \\
   + \left[\frac{1-p}{2} \Pr(U_\ell \in A_{\ell,+}|Y_\ell = -1) +
        p \Pr(U_\ell \in A_{\ell,+}|Y_\ell = 0) \right].
\end{multline*}
Since $U_\ell \markov Y_\ell \markov Y_0$,
\begin{align*}
\Pr(\tilde{U}_\ell = 5)
  & \le \Bigg[\frac{1-p}{2} \Pr(U_\ell \in A_{\ell,-}|Y_\ell = 1,Y_0 = 1) \\
  & \phantom{=} + p \Pr(U_\ell \in A_{\ell,-}|Y_\ell = 0,Y_0 = 1)\Bigg] \\
  &\phantom{=}  + \Bigg[\frac{1-p}{2} 
           \Pr(U_\ell \in A_{\ell,+}|Y_\ell = -1,Y_0 = -1) \\
  & \phantom{=} + p \Pr(U_\ell \in A_{\ell,+}|Y_\ell = 0,Y_0 = -1) \Bigg] \\
  & \le \Pr(U_\ell \in A_{\ell,-}|Y_0 = 1) + 
      \Pr(U_\ell \in A_{\ell,+}|Y_0 = -1).
\end{align*}
But
\begin{equation*}
\sum_{\mathbf{u} : \phi(\mathbf{u}) = 1} \frac{1}{2} \prod_{j = 1}^L
   \Pr(U_j = u_j|Y_0 = -1) \lambda \le D,
\end{equation*}
which implies
\begin{equation}
\label{preHolder}
\sum_{\mathbf{u} : \phi(\mathbf{u}) = 1, u_\ell \in A_{\ell,+}} 
    \prod_{j = 1}^L \Pr(U_j = u_j|Y_0 = -1) \le \frac{2D}{\lambda}.
\end{equation}
By the definition of $A_{\ell,+}$, for each $u_\ell \in A_{\ell,+}$,
there exists at least one 
$\mathbf{u}_{\ell^c}$ such that $\phi(\mathbf{u}) = 1$
and 
\begin{equation*}
\prod_{j = 1}^L \Pr(U_j = u_j|Y_0 = -1) \ge 
   \Pr(U_\ell = u_\ell|Y_0 = -1)^L.
\end{equation*}
Together with~(\ref{preHolder}), this implies
\begin{equation*}
\sum_{u_\ell \in A_{\ell,+}} 
    \Pr(U_\ell = u_\ell|Y_0 = -1)^L \le \frac{2D}{\lambda}.
\end{equation*}
Applying H\"{o}lder's inequality~\cite[p.~121]{Royden:Real} gives
\begin{align*}
\Pr(U_\ell \in A_{\ell,+}|Y_0 = -1) & = \sum_{u_\ell \in A_{\ell,+}}
    \Pr(U_\ell = u_\ell|Y_0 = -1)  \\
  & \le \left[\sum_{u_\ell \in A_{\ell,+}}
   \Pr(U_\ell = u_\ell|Y_0 = -1)^L\right]^{1/L} \cdot |A_{\ell,+}|^{(L-1)/L} \\
   & \le 4 \left(\frac{2D}{\lambda}\right)^{1/L}.
\end{align*}
Likewise,
\begin{equation*}
\Pr(U_\ell \in A_{\ell,-}|Y_0 = 1) \le 4 \left(\frac{2D}{\lambda}\right)^{1/L}.
\end{equation*}
Thus
\begin{equation*}
\Pr(\tilde{U}_\ell = 5) \le 8 \left(\frac{2D}{\lambda}\right)^{1/L}.
\end{equation*}
By the union bound, it follows that
\begin{equation*}
\Pr(\max(\tilde{U}_1,\ldots,\tilde{U}_L) = 5) \le 
  8L\left(\frac{2D}{\lambda}\right)^{1/L}
\end{equation*}
and therefore
\begin{equation*}
E[d_1^\lambda(Y_0,\tilde{Z}_1)] \le D + 
  8L\left(\frac{2D}{\lambda}\right)^{1/L} \le D + \delta.
\end{equation*}
Note that $\tilde{Z}_1 = 1$ only if $\tilde{U}_\ell$ is in
$A_{\ell,+}$ for some $\ell$, and 
\begin{equation*}
\Pr(\tilde{U}_\ell \in A_{\ell,+}|Y_0 = -1) = 0 \ \text{for all $\ell$}. 
\end{equation*}
Thus $\Pr(Y_0 = -1,\tilde{Z}_1 = 1) = 0$ and
similarly,
$\Pr(Y_0 = 1,\tilde{Z}_1 = -1) = 0$.
It follows from Lemma~\ref{central} that
\begin{equation}
\label{midway}
\frac{1}{L} \sum_{\ell = 1}^L I(Y_\ell;\tilde{U}_\ell|Y_0) 
   \ge g\left((D + \delta)^{1/L}\right).
\end{equation}
The remainder of the proof is devoted to showing that 
$I(Y_\ell;\tilde{U}_\ell|Y_0)$ is close to $I(Y_\ell;U_\ell|Y_0)$.
For this we use the decomposition
\begin{equation*}
I(Y_\ell;\tilde{U}_\ell|Y_0) = H(\tilde{U}_\ell|Y_0) - H(\tilde{U}_\ell|Y_\ell).
\end{equation*}
Observe that
\begin{equation*}
\Pr(\tilde{U}_\ell = 5|Y_\ell = 0) p \le \Pr(\tilde{U}_\ell = 5)
  \le 8\left(\frac{2D}{\lambda}\right)^{1/L}.
\end{equation*}
Thus
\begin{equation*}
\Pr(\tilde{U}_\ell = 5|Y_\ell = 0) \le \frac{\delta}{2}.
\end{equation*}
Similarly,
\begin{align*}
\Pr(\tilde{U}_\ell = 5|Y_\ell = 1) & \le \frac{\delta}{2} \\
\intertext{and}
\Pr(\tilde{U}_\ell = 5|Y_\ell = -1) & \le \frac{\delta}{2}.
\end{align*}
Therefore if we view $U_\ell$ as a random variable on $\{1,\ldots,5\}$,
for any $i$ in $\{-1,0,1\}$,
\begin{equation*}
\sum_{j = 1}^5 |\Pr(\tilde{U}_\ell = j|Y_\ell = i) - 
   \Pr(U_\ell = j|Y_\ell = i)| = 2\Pr(\tilde{U}_\ell = 5|Y_\ell = i) \le \delta.
\end{equation*}
A standard result on the continuity of entropy~\cite[Lemma~1.2.7]{CK:IT}
now implies that (recall $\delta \le 1/2$)
\begin{equation*}
|H(\tilde{U}_\ell|Y_\ell = i) - H(U_\ell|Y_\ell = i)| \le - 
     \delta \log \frac{\delta}{5}
\end{equation*}
so
\begin{equation*}
|H(\tilde{U}_\ell|Y_\ell) - H(U_\ell|Y_\ell)| \le - 
      \delta \log \frac{\delta}{5}.
\end{equation*}
Likewise, for any $i$ in $\{-1,1\}$,
\begin{equation*}
\frac{1}{2} \Pr(\tilde{U}_\ell = 5|Y_0 = i) \le 
  8\left(\frac{2D}{\lambda}\right)^{1/L}.
\end{equation*}
Thus
\begin{equation*}
\Pr(\tilde{U}_\ell = 5|Y_0 = i) \le \frac{\delta}{2},
\end{equation*}
so
\begin{equation*}
|H(\tilde{U}_\ell|Y_0) - H(U_\ell|Y_0)| \le - \delta \log \frac{\delta}{5}
\end{equation*}
as before. It follows that
\begin{equation*}
|I(Y_\ell;\tilde{U}_\ell|Y_0) - I(Y_\ell;U_\ell|Y_0)| \le 
   - 2 \delta \log \frac{\delta}{5}.
\end{equation*}
Combining this with~(\ref{midway}) yields
\begin{equation*}
\frac{1}{L} \sum_{\ell = 1}^L I(Y_\ell;U_\ell|Y_0) \ge 
   g\left((D + \delta)^{1/L}\right) +
  2\delta \log \frac{\delta}{5}. 
\end{equation*}
\QED

We are now in a position to prove the main result of
this Appendix.
\begin{lemma} For any $p^L \le D$,
\begin{equation*}
\lim_{\lambda \rightarrow \infty} \mathcal{R}_o(D,\lambda) 
   \ge (1 - D) \log 2 + L g\left(D^{1/L}\right).
\end{equation*}
\end{lemma}

\emph{Proof.} 
Fix $p^L \le D \le 1$ and $\delta \in (0,1/2]$, and suppose
$\lambda$ satisfies
\begin{equation}
\label{biglambda}
\lambda \ge \max\left[4\left(\frac{32 L}{\delta p (1-p)}\right)^{2L},
   \left(\frac{D}{\delta}\right)^2\right].
\end{equation}
By taking $X = Y_0$ in the definition of $\mathcal{RD}_o(\lambda)$,
it follows that there exist $\mathbf{R}$ in $\mathbb{R}^L_+$ and
$\gamma$ in $\Gamma_o$ such that
\begin{equation}
\label{sumrate}
\begin{split}
D + \delta & \ge E[d_1^\lambda(Y_0,Z_1)], \ \text{and} \\
\mathcal{R}_o(D,\lambda) + \delta \ge 
\sum_{\ell = 1}^L R_\ell & \ge I(Y_0;\mathbf{U}|T) +
 \sum_{\ell = 1}^L I(Y_\ell;U_\ell|Y_0,W,T).
\end{split}
\end{equation}
For each possible realization $(w,t)$ of $(W,T)$, let
\begin{equation*}
D_{w,t} = E[d_1^\lambda(Y_0,Z_1)|W = w, T = t].
\end{equation*}
Let $S = \{(w,t) : D_{w,t} \le \sqrt{\lambda}\}$.
Then by Markov's inequality,
\begin{equation}
\label{markov}
\Pr((W,T) \notin S) \le \frac{D}{\sqrt{\lambda}} \le \delta.
\end{equation}
In particular, $\Pr((W,T) \in S) > 0$.
Also, for any $(w,t) \in S$,
\begin{equation*}
\frac{32 L}{p(1-p)} \left(\frac{2 D_{w,t}}{\lambda}\right)^{1/L}
\le \delta
\end{equation*}
by~(\ref{biglambda}).
Thus, by Lemma~\ref{continuity}, if $(w,t) \in S$,
\begin{equation*}
\frac{1}{L} \sum_{\ell = 1}^L I(Y_\ell;U_\ell|Y_0, W = w, T = t) \ge
 g\left((D_{w,t} + \delta)^{1/L}\right) + 2 \delta \log \frac{\delta}{5}.
\end{equation*}
By averaging over $(w,t) \in S$ and 
invoking Corollary~\ref{convexcor}, we obtain
\begin{multline*}
\sum_{(w,t) \in S}
\frac{1}{L} \sum_{\ell = 1}^L I(Y_\ell;U_\ell|Y_0, W = w, T = t) \cdot
\frac{\Pr(W = w, T = t)}{\Pr((W,T) \in S)} \ge \\
 g((D + \delta)^{1/L}) + 2 \delta \log \frac{\delta}{5}.
\end{multline*}
From~(\ref{markov}), it follows that
\begin{equation}
\label{sumnoise}
\sum_{\ell = 1}^L I(Y_\ell;U_\ell|Y_0,W,T) \ge L(1-\delta) \left[
  g\left((D+\delta)^{1/L}\right) + 2\delta \log \frac{\delta}{5}\right].
\end{equation}
Now by the data processing inequality,
\begin{align*}
I(Y_0;\mathbf{U}|T) & = I(Y_0;\mathbf{U},T) \\
  & \ge I(Y_0;Z_1).
\end{align*}
Let $\varepsilon = 1(Y_0\cdot Z_1 = -1)$. Continuing,
\begin{align*}
I(Y_0;\mathbf{U}|T) & \ge H(Y_0) - H(Y_0|Z_1) \\
       & =\log 2 - H(Y_0,\varepsilon|Z_1) \\
       & = \log 2 - H(\varepsilon|Z_1) - H(Y_0|\varepsilon,Z_1) \\
       & \ge \log 2 - h(D/\lambda) - \Pr(Z_1 = 0) \log 2 \\
       & \ge (1 -D)\log 2 - h(\delta).
\end{align*}
Substituting this and~(\ref{sumnoise}) into~(\ref{sumrate}) yields
\begin{equation*}
R_o(D,\lambda) 
  \ge (1-D)\log 2 - h(\delta) + L(1-\delta)\left[
  g\left((D + \delta)^{1/L}\right) + 
   2\delta \log \frac{\delta}{5}\right] - \delta.
\end{equation*}
The proof is terminated by letting $\lambda \rightarrow \infty$
and then $\delta \rightarrow 0$. \QED

\section{The Berger-Tung Outer Bound is Loose for the Binary Erasure
CEO Problem}
\label{BT:loose:BE}

We will show numerically that for one instance of the binary erasure 
CEO problem,
$\mathcal{RD}_o^{BT}$ contains points with a strictly superoptimal
sum rate. Let $L = 2$ and $p = 1/2$. Let $W_1$ and $W_2$ be $\{0,1\}$-valued
random variables with the joint distribution
\begin{equation*}
\left[
\begin{array}{cc}
1/5 & 2/5 \\
2/5 & 0
\end{array}
\right],
\end{equation*}
i.e.,
\begin{align*}
\Pr(W_1 = 0, W_2 = 0) & = \frac{1}{5} \\
\Pr(W_1 = 1, W_2 = 0) & = \Pr(W_1 = 0, W_2 = 1) = \frac{2}{5}.
\end{align*}
We assume that $(W_1,W_2)$ is independent of ($Y_0,Y_1,Y_2)$.
Let $U_\ell = Y_\ell \cdot W_\ell$ for $\ell$ in 
$\{1,2\}$, and let $Z_1 = \sgn(U_1 + U_2)$. Since
$Y_\ell$ can be written as $Y_\ell = Y_0 \cdot N_\ell$ where $N_1$ and
$N_2$ are i.i.d. with $\Pr(N_1 = 0) = \Pr(N_1 = 1) = 1/2$
(recall the notation of Section~\ref{BECEO:subsection}), we have
$U_\ell = Y_0 \cdot N_\ell \cdot W_\ell$. Note that $N_1 \cdot W_1$ and
$N_2 \cdot W_2$ have the joint distribution
\begin{equation*}
\left[
\begin{array}{cc}
3/5 & 1/5 \\
1/5 & 0   
\end{array}
\right].
\end{equation*}
Thus for any $\lambda$,
$E[d_1^\lambda(Y_0,Z_1)] = \Pr(N_1 \cdot W_1 = N_2 \cdot W_2 = 0) = 3/5$.
Now we can compute
\begin{equation*}
I(Y_1,Y_2;U_1,U_2) \le 0.6273 \ \text{nats}
\end{equation*}
and
\begin{align*}
I(Y_1,Y_2;U_1|U_2) & = I(Y_1,Y_2;U_1,U_2) - I(Y_1,Y_2;U_2) \\
         & = I(Y_1,Y_2;U_1,U_2) - I(Y_2;U_2) \\
         & \le 0.3248 \ \text{nats}.
\end{align*}
It follows that $(0.3248, 0.3248, 3/5)$ is in $\mathcal{RD}_o^{BT}(\lambda)$ 
for any $\lambda$. Thus
\begin{equation*}
\lim_{\lambda \rightarrow \infty} \inf\left\{R_1 + R_2 : 
  \left(R_1,R_2,\frac{3}{5}\right)
  \in \overline{\mathcal{RD}_o^{BT}}(\lambda) \right\} 
  \le 0.6496 \ \text{nats}.
\end{equation*}
From the previous two appendices,
the correct sum rate is
\begin{equation*}
\left(1-\frac{3}{5}\right)\log 2 + 
   2g\left(\sqrt{\frac{3}{5}}\right) \ge 0.6562 \ \text{nats}.
\end{equation*}

\section{Evaluation of the Outer Bound for the Gaussian CEO Problem}
\label{Gaussian:proof}
Two lemmas are needed for our proof of Proposition~\ref{Gaussian:Prop}.
The first is a simple extension of Theorem~\ref{main} to the
Gaussian CEO problem setting of Section~\ref{CEO:Gaussian}.
For this appendix, let us redefine $\chi$ to be
the set of real-valued random variables $X$ such that $Y_1,\ldots,Y_L$
are conditionally independent given $X$ (the side information
$Y_{L+1}$ is unneeded in this context and shall be ignored). 
Let us also redefine $\Gamma_o$
to be the set of random variables $(U_1,\ldots,U_L,Z_1,W,T)$
such that each takes values in a finite-dimensional Euclidean space,
and collectively they satisfy the Markov conditions defining the
original $\Gamma_o$,
\begin{enumerate}
\item[(\emph{i})]
$(W,T)$ is independent of $(Y_0,\mathbf{Y})$,
\item[(\emph{ii})]
$U_\ell \markov (Y_\ell,W,T) \markov 
   (Y_0,\mathbf{Y}_{\ell^c},\mathbf{U}_{\ell^c})$ for
 all $\ell$, and
\item[(\emph{iii})]
$(Y_0,\mathbf{Y},W) \markov (\mathbf{U},T) \markov Z_1$,
\end{enumerate}
and one new technical condition,
\begin{enumerate}
\item[(\emph{iv})] the conditional distribution of $U_\ell$ given $W$
and $T$ is discrete for each $\ell$.
\end{enumerate}
Note that any conditional distribution involving these random
variables is well-defined~\cite[Theorem~6.3]{Kallenberg:Foundations:2}.
As such, so is any conditional mutual information~\cite[Ch.~3, 
especially the translator's notes at the end]{Pinsker:Stability}.
\begin{lemma}
For the Gaussian CEO problem, $\mathcal{RD}_\star \subset
\mathcal{RD}_o$ if $\mathcal{RD}_o$ is defined using the
$\chi$ and $\Gamma_o$ just described.
\end{lemma}

The proof follows the original and is omitted.
The second ingredient is a consequence of an intriguing result
of Oohama~\cite{Oohama:CEO:Region} and 
Prabhakaran, Tse, and Ramchandran that relates
information the encoders send about the hidden source
to information they send about their observation
``noise.'' 
\begin{lemma}[c.f.~{\cite[Lemma~3]{Oohama:CEO:Region}}]
\label{core}
If $\gamma$ is in
$\Gamma_o$, then for all $A \subset \{1,\ldots,L\}$,
\begin{equation*}
\exp(2I(Y_0;\mathbf{U}_A|W,T)) \le
1 + \sum_{\ell \in A} \frac{1 - \exp(-2I(Y_\ell;U_\ell|Y_0,W,T))}
  {\sigma_\ell^2/\sigma^2}.
\end{equation*}
\end{lemma}
\emph{Proof.} 
For any realization of $(W,T)$, it follows from Lemma~3 in
Oohama~\cite{Oohama:CEO:Region} that\footnote{Oohama's result
assumes that $U_\ell$ is a discrete, deterministic function of
$Y_\ell$ for each $\ell$, but the proof shows that conditions
(\emph{ii}) and (\emph{iv}) above are actually sufficient.} 
\begin{multline*}
\exp(2I(Y_0;\mathbf{U}_A|W = w, T = t)) \le \\
  1 + \sum_{\ell \in A} \frac{1 - \exp(-2I(Y_\ell;U_\ell|Y_0,W = w, T = t))}%
  {\sigma_\ell^2/\sigma^2}.
\end{multline*}
We now average over $(w,t)$ and invoke the convexity of $\exp(\cdot)$
twice, once on each side. \QED

\emph{Proof of Proposition~\ref{Gaussian:Prop}}. 
If $(\mathbf{R},D)$ is in $\mathcal{RD}_o$, then there exists
$\gamma$ in $\Gamma_o$ such that 
$E[(Y_0 - Z_1)^2] \le D$ and for all $A \subset \{1,\ldots,L\}$,
\begin{equation}
\label{Gaussian:proof:var}
\sum_{\ell \in A} R_\ell \ge I(Y_0;\mathbf{U}_A|\mathbf{U}_{A^c},T) + 
   \sum_{\ell \in A} I(Y_\ell;U_\ell|Y_0,W,T).
\end{equation}
Now
\begin{equation}
\label{Gaussian:main:formula}
I(Y_0;\mathbf{U}_A|\mathbf{U}_{A^c},T) + I(Y_0;\mathbf{U}_{A^c}|T) =
  I(Y_0;\mathbf{U}|T).
\end{equation}
Since $Y_0 \markov (\mathbf{U},T) \markov Z_1$, the
right-hand side can be lower bounded as follows
\begin{align*}
I(Y_0;\mathbf{U}|T) 
          & =I(Y_0;\mathbf{U},T) \\
          & \ge I(Y_0;Z_1) \\
          & \ge \frac{1}{2}\log^+ \frac{\sigma^2}{E[(Y_0 - Z_1)^2]},
\end{align*}
where we have used the rate-distortion theorem for Gaussian
sources~\cite[Theorem~13.3.2]{Cover:IT}. 
In particular,
\begin{equation}
\label{Gaussian:RD}
I(Y_0;\mathbf{U}|T) \ge \frac{1}{2} \log \frac{\sigma^2}{D}.
\end{equation}
Let us address the second term on the left-hand side 
of~(\ref{Gaussian:main:formula}). Observe that
\begin{align*}
I(Y_0;\mathbf{U}_{A^c}|T) + I(Y_0;W|\mathbf{U}_{A^c},T)
  & = I(Y_0;W|T) + I(Y_0;\mathbf{U}_{A^c}|W,T) \\
  & = I(Y_0;\mathbf{U}_{A^c}|W,T). 
\end{align*}
Defining $r_\ell = I(Y_\ell;U_\ell|Y_0,W,T)$ and applying
Lemma~\ref{core} to the right-hand side gives
\begin{equation}
\label{invokingkey}
I(Y_0;\mathbf{U}_{A^c}|T) \le
  \frac{1}{2} \log\left[1 + \sum_{\ell \in A^c} 
   \frac{1 - \exp(-2r_\ell)}{\sigma^2_\ell/\sigma^2}\right].
\end{equation}
Substituting (\ref{Gaussian:RD}) and
(\ref{invokingkey}) into (\ref{Gaussian:main:formula}) gives
\begin{equation*}
I(Y_0;\mathbf{U}_A|\mathbf{U}_{A^c},T) \ge
 \frac{1}{2} \log^+ \left\{ \frac{1}{D} \left[\frac{1}{\sigma^2} + 
   \sum_{\ell \in A^c} \frac{1 - \exp(-2r_\ell)}{\sigma^2_\ell} \right]^{-1}
  \right\}.
\end{equation*}
The conclusion follows upon substitution of this inequality and the
definition of $r_\ell$ into~(\ref{Gaussian:proof:var}). \QED

\section{The Berger-Tung Outer Bound is Loose for the Gaussian CEO Problem}
\label{BT:loose:Gaussian}

We have just seen that the improved outer
bound is capable of recovering the converse result of
Oohama~\cite{Oohama:CEO:Region} and Prabhakaran, Tse, and
Ramchandran~\cite{Prabhakaran:ISIT04} for the Gaussian
CEO problem.
Here we will show that the Berger-Tung outer bound does not
recover this result.
As with the binary erasure CEO problem, we will show that,
in general, the
Berger-Tung outer bound contains points with a strictly 
superoptimal sum rate.

Consider the case in which, in the notation of 
Section~\ref{CEO:Gaussian},
$L = 2$ and $\sigma^2 = \sigma_1^2 = \sigma_2^2 = 1$.
In words, two encoders each observe a unit variance, i.i.d.\
Gaussian process in additive Gaussian noise
with a signal-to-noise ratio of unity. It follows from
Proposition~\ref{Gaussian:Prop}
that the minimum sum rate needed to achieve 
the distortion $1/2$ is at least $(3/2)\log 2$ nats.

Let $W$, $V_1$, and $V_2$ be Gaussian random variables,
independent of each other and $Y_0$, $Y_1$, and $Y_2$.
Let $V_1$ and $V_2$ have unit variance; we denote
the variance of $W$ by $\sigma^2_W$. Let
\begin{align*}
U_1 & = Y_1 + V_1 + W \\
U_2 & = Y_2 + V_2 - W.
\end{align*}
Note that the sum of $U_1$ and $U_2$ is a sufficient statistic for
$Y_0$ given $U_1$ and $U_2$. This observation makes it easy to
verify that if $Z_1 = E[Y_0|U_1,U_2]$, then
$E[(Y_0 - Z_1)^2] = 1/2$. Note that this distortion is independent of 
$\sigma^2_W$.

It follows that for any value of $\sigma^2_W$, $\mathcal{RD}_o^{BT}$
contains points of the form $(R_1,R_2,1/2)$ with
\begin{equation*}
R_1 + R_2 = \max(I(Y_1,Y_2;U_1,U_2),2 \; I(Y_1,Y_2;U_1|U_2)).
\end{equation*}
But~\cite[Theorem~9.4.1]{Cover:IT}
\begin{equation*}
I(Y_1,Y_2;U_1,U_2) = \frac{3}{2}\log 2 + \frac{1}{2} \log 
   \frac{1 + \sigma^2_W}{1 + 2 \sigma^2_W}
\end{equation*}
and
\begin{align*}
I(Y_1,Y_2;U_1|U_2) & = I(Y_1,Y_2;U_1,U_2) - I(Y_1,Y_2;U_2) \\
      & = I(Y_1,Y_2;U_1,U_2) - I(Y_2;U_2) \\
      & = \frac{3}{2}\log 2 + \frac{1}{2} \log 
    \frac{1 + \sigma^2_W}{1 + 2 \sigma^2_W} - \frac{1}{2}\log\left(1 +
   \frac{2}{\sigma^2_W +1}\right).
\end{align*}
Observe that, when viewed as functions of $\sigma^2_W$, $I(Y_1,Y_2;U_1,U_2)$
is strictly decreasing and $I(Y_1,Y_2;U_1|U_2)$ is continuous. 
Since $\sigma_W^2 = 0$ yields 
\begin{align*}
I(Y_1,Y_2;U_1,U_2) & = \frac{3}{2} \log 2 \\
2I(Y_1,Y_2;U_1|U_2) & < \frac{3}{2} \log 2,
\end{align*}
it follows that there exists $\sigma^2_W > 0$ such that
\begin{equation*}
\max(I(Y_1,Y_2;U_1,U_2),2I(Y_1,Y_2;U_1|U_2)) < \frac{3}{2} \log 2.
\end{equation*}

\section{$\mathcal{RD}_i^{BT}$ is Closed}
\label{closed}

The main step in proving that $\mathcal{RD}_i^{BT}$
is closed is to show that one can limit the ranges of the
auxiliary random variables without reducing the region.

\begin{defn}
\label{BT:hat:defn}
Let $\hat{\Gamma}_i^{BT}$ denote the set of finite-alphabet random
variables 
\begin{equation*}
\gamma = (U_1,\ldots,U_L, Z_1,\ldots,Z_K,T)
\end{equation*}
in $\Gamma_i^{BT}$ such that 
\begin{equation*}
|\mathcal{U}_\ell| = |\mathcal{Y}_\ell| + 2^L + K - 1 \ \text{for all $\ell$}
\end{equation*}
and 
\begin{equation*}
|\mathcal{T}| = 2^L + K.
\end{equation*}
Then let
\begin{equation*}
\hat{\mathcal{RD}}_i^{BT} = \bigcup_{\gamma \in \hat{\Gamma}_i^{BT}}
\mathcal{RD}_i^{BT}(\gamma).
\end{equation*}
\end{defn}

We shall show that $\mathcal{RD}_i^{BT} = \hat{\mathcal{RD}}_i^{BT}$
in two steps, first handling the case in which $T$ is deterministic,
and then bootstrapping to the general case. Both steps involve
now-standard uses of Carath\'{e}odory's
theorem~\cite[Theorem~17.1]{Rockafellar:Convex}. 
We give proofs of both steps, albeit condensed ones,
due to the complexity of our setup.
\begin{lemma}
\label{deterministic}
Suppose that $\gamma = (\mathbf{U},\mathbf{Z},T)$ 
is in $\mathcal{RD}_i^{BT}$
and $T$ is deterministic. Then there exists 
$\hat{\gamma} = (\hat{\mathbf{U}},\hat{\mathbf{Z}},\hat{T})$ in 
$\hat{\mathcal{RD}}_i^{BT}$ such that
$\hat{T}$ is deterministic and 
$\mathcal{RD}_i^{BT}(\gamma) =
  \mathcal{RD}_i^{BT}(\hat{\gamma})$.
\end{lemma}
\emph{Proof.} 
For any $A \subset \{1,\ldots,L\}$ containing $1$, we have
\begin{align*}
I(\mathbf{Y}_A;\mathbf{U}_A|\mathbf{U}_{A^c}) & = 
             H(\mathbf{Y}_A|\mathbf{U}_{A^c}) - 
               H(\mathbf{Y}_A|\mathbf{U}) \\
        & = H(\mathbf{Y}_A|\mathbf{U}_{A^c}) - 
                \sum_{u_1} H(\mathbf{Y}_A|\mathbf{U}_{1^c},U_1 = u_1) 
                 \Pr(U_1 = u_1),
\end{align*}
while for any nonempty $A$ not containing $1$, we have
\begin{equation*}
I(\mathbf{Y}_A;\mathbf{U}_A|\mathbf{U}_{A^c}) = \sum_{u_1}
       I(\mathbf{Y}_A;\mathbf{U}_A|\mathbf{U}_{A^c\backslash\{1\}},U_1 = u_1)
    \Pr(U_1 = u_1).
\end{equation*}
Carath\'{e}odory's theorem guarantees that we can find a 
$\hat{U}_1$ with $\hat{\mathcal{U}}_1 \subset \mathcal{U}_1$ such
that $|\hat{\mathcal{U}}_1| = |\mathcal{Y}_1| + 2^L + K - 1$,
\begin{equation*}
\sum_{u_1 \in \hat{\mathcal{U}}_1} \Pr(Y_1 = y_1|U_1 = u_1) 
   \Pr(\hat{U}_1 = u_1) =
   \Pr(Y_1 = y_1) \ \text{for all $y_1$ in
   $\mathcal{Y}_1$ but one},
\end{equation*}
\begin{multline*}
\sum_{u_1 \in \mathcal{U}_1} H(\mathbf{Y}_A|\mathbf{U}_{1^c},U_1 = u_1) 
      \Pr(U_1 = u_1) =  \\
   \sum_{u_1 \in \hat{\mathcal{U}}_1} 
   H(\mathbf{Y}_A|\mathbf{U}_{1^c},U_1 = u_1) 
      \Pr(\hat{U}_1 = u_1) \ \text{for all $A$ containing 1},
\end{multline*}
and similarly for
    $I(\mathbf{Y}_A;\mathbf{U}_A|\mathbf{U}_{A^c\backslash \{1\}},U_1 = u_1)$ 
and
$E[d_k(Y_0,\mathbf{Y},Y_{L+1},Z_k)|U_1 = u_1]$.
Since $U_1 \markov Y_1 \markov (Y_0,\mathbf{Y}_{1^c},\mathbf{U}_{1^c},
Y_{L+1})$, if we substitute $\hat{U}_1$ for $U_1$, the resulting
$\gamma$ is in $\Gamma_i^{BT}$ and 
$\mathcal{RD}_i^{BT}(\gamma)$ is unchanged. Repeating this
procedure for $U_2,\ldots,U_L$ completes the proof.  \QED
\begin{lemma}
$\mathcal{RD}_i^{BT} = \hat{\mathcal{RD}}_i^{BT}$.
\end{lemma}
\emph{Proof.}
Let $(\mathbf{U},\mathbf{Z},T)$ be in $\Gamma_i^{BT}$. For each
$t$ in $\mathcal{T}$, let $(\mathbf{U},\mathbf{Z},t)$ denote
the joint distribution of $(\mathbf{U},\mathbf{Z},T)$ conditioned
on the event $\{T = t\}$. By Lemma~\ref{deterministic}, for each $t$,
there exists $(\hat{\mathbf{U}},\hat{\mathbf{Z}})$ such that
$(\hat{\mathbf{U}},\hat{\mathbf{Z}},t)$ is in $\hat{\Gamma}_i^{BT}$ and
$\mathcal{RD}_i^{BT}(\mathbf{U},\mathbf{Z},t) =
\mathcal{RD}_i^{BT}(\hat{\mathbf{U}},\hat{\mathbf{Z}},t)$.
By replacing $(\mathbf{U},\mathbf{Z})$ with 
$(\hat{\mathbf{U}},\hat{\mathbf{Z}})$ for each value of $T$, we
obtain $(\hat{\mathbf{U}},\hat{\mathbf{Z}},T)$ in 
$\Gamma_i^{BT}$ such that 
$|\hat{\mathcal{U}}_\ell| = |\mathcal{Y}_\ell| + 2^L + K - 1$ for all $\ell$
and $\mathcal{RD}_i^{BT}(\mathbf{U},\mathbf{Z},T) =
\mathcal{RD}_i^{BT}(\hat{\mathbf{U}},\hat{\mathbf{Z}},T)$. Now
\begin{multline*}
\mathcal{RD}_i^{BT}(\hat{\mathbf{U}},\hat{\mathbf{Z}},T) 
         = \Bigg\{ (\mathbf{R},\mathbf{D}) : \sum_{\ell \in A} R_\ell \ge \\
   \sum_{t \in \mathcal{T}} I(\mathbf{Y}_A;\mathbf{U}_A|%
       \mathbf{U}_{A^c},Y_{L+1},T = t)\Pr(T = t) \ \text{for all $A$, and} \\
D_k \ge \sum_{t \in \mathcal{T}} 
   E[d_k(Y_0,\mathbf{Y},Y_{L+1},Z_k)|T = t]\Pr(T = t)
    \ \text{for all $k$} \Bigg\}.
\end{multline*}
Carath\'{e}odory's theorem
implies that we can find a $\hat{T}$ with
$\hat{\mathcal{T}} \subset \mathcal{T}$ and $|\hat{\mathcal{T}}| =
2^L + K$ such that
\begin{multline*}
\sum_{t \in \mathcal{T}} I(\mathbf{Y}_A;\mathbf{U}_A|%
       \mathbf{U}_{A^c},Y_{L+1},T = t)\Pr(T = t) \\
 = \sum_{t \in \hat{\mathcal{T}}} I(\mathbf{Y}_A;\mathbf{U}_A|%
       \mathbf{U}_{A^c},Y_{L+1},T = t)\Pr(\hat{T} = t) \
          \text{for all $A$} 
\end{multline*}
and similarly for $E[d_k(Y_0,\mathbf{Y},Y_{L+1},Z_k)|T = t]$.
Then $(\hat{\mathbf{U}},\hat{\mathbf{Z}},\hat{T})$ is
in $\hat{\Gamma}_i^{BT}$ and 
\begin{equation*}
\mathcal{RD}_i^{BT}(\mathbf{U},\mathbf{Z},T) 
= \mathcal{RD}_i^{BT}(\hat{\mathbf{U}},
    \hat{\mathbf{Z}},\hat{T}). 
\end{equation*}
Since $(\mathbf{U},\mathbf{Z},T)$ in
$\Gamma_i^{BT}$ was arbitrary, it holds 
$\mathcal{RD}_i^{BT} \subset \hat{\mathcal{RD}}_i^{BT}$. This
completes the proof since the reverse containment is obvious. \QED

The cardinality bounds provided by the last two lemmas,
while finite, are exponential in $L$ and hence impractical for
moderate numbers of encoders. One can improve upon these
bounds by exploiting the polymatroid 
structure~\cite{Viswanath:DIMACS:Source,Chen:CEO:Bounds}
of $\mathcal{RD}_i^{BT}$. While this would be useful if one wished
to numerically evaluate the bound, our aim here is merely to show
that it is closed.
\begin{lemma}
$\hat{\mathcal{RD}}_i^{BT}$ is closed.
\end{lemma}
\emph{Proof.} 
The Markov conditions defining $\hat{\Gamma}_i^{BT}$ can be expressed as 
\begin{align*}
I(T;Y_0,\mathbf{Y},Y_{L+1}) & = 0 \\
I(U_\ell;Y_0,\mathbf{Y}_{\ell^c},Y_{L+1},\mathbf{U}_{\ell^c}|Y_\ell,T) & = 0 
 \ \text{for all $\ell$} \\
I(Y_0,\mathbf{Y};\mathbf{Z}|\mathbf{U},Y_{L+1},T) & = 0.
\end{align*}
Since the conditional mutual information function is continuous,
$\hat{\Gamma}_i^{BT}$ is compact when viewed as a subset of Euclidean
space. Thus if $(\mathbf{R}^{(n)},\mathbf{D}^{(n)})$ is a sequence in
$\hat{\mathcal{RD}}_i^{BT}$ that converges to $(\mathbf{R},\mathbf{D})$,
by considering subsequences we may assume that 
$(\mathbf{R}^{(n)},\mathbf{D}^{(n)})$
is in $\hat{\mathcal{RD}}_i^{BT}(\gamma^{(n)})$ for each $n$ and
$\gamma^{(n)} \rightarrow \gamma = (\mathbf{U},\mathbf{Z},T) 
\in \hat{\Gamma}_i^{BT}$.
By invoking the continuity of mutual information once again, we obtain
\begin{equation*}
\sum_{\ell \in A} R_\ell \ge I(\mathbf{Y}_A;\mathbf{U}_A|\mathbf{U}_{A^c},T)
\end{equation*}
for each $A$. Likewise,
\begin{equation*}
D_k \ge E[d_k(Y_0,\mathbf{Y},Y_{L+1},Z_k)] \ \text{for all $k$}.
\end{equation*}
It follows that $(\mathbf{R},\mathbf{D})$ 
is in $\mathcal{RD}_i^{BT}(\gamma)$ and
therefore also in $\hat{\mathcal{RD}}_i^{BT}$. \QED
\begin{cor}
$\mathcal{RD}_i^{BT}$ is closed.
\end{cor}

\section*{Acknowledgment}

It is a pleasure to acknowledge
discussions with Vinod Prabhakaran.
The results in 
Section~\ref{CEO:Gaussian} are due to him. This
work has also benefited from the helpful comments of
Stark~C. Draper, Pramod~Viswanath, and Anant Sahai.

\end{document}